\documentclass[11pt,letterpaper]{article}
\pdfoutput=1
\usepackage{jheppub}
\usepackage{color}
\usepackage{graphicx}
\usepackage{wrapfig}

\usepackage{verbatim}
\usepackage{amsmath}
\usepackage{amssymb}
\usepackage{subfig}
\usepackage{url}
\usepackage{bbold}
\usepackage{xspace}

\usepackage{multirow}
\usepackage{threeparttable}
\usepackage{paralist}

\usepackage{color}

\usepackage{floatrow}
\newfloatcommand{capbtabbox}{table}[][\FBwidth]

\usepackage{tikz}
\usetikzlibrary{trees}
\usetikzlibrary{decorations.pathmorphing}
\usetikzlibrary{decorations.markings}
\usetikzlibrary{arrows}

\definecolor{darkgreen}{rgb}{0,0.5,0}

\DeclareRobustCommand{\Sec}[1]{Sec.~\ref{#1}}

\DeclareRobustCommand{\Fig}[1]{Fig.~\ref{#1}}

\newcommand{\Dfbd}{\mathord{\buildrel{\lower3pt\hbox{$\scriptscriptstyle\leftrightarrow$}}\over {D}_{\mu}}}

\newcommand{\beq}{\begin{equation}}
\newcommand{\eeq}[1]{\label{#1}\end{equation}}
\def\beqa{\begin{eqnarray}}
\def\eeqa#1{\label{#1}\end{eqnarray}}
\newcommand{\eeqn}{\end{equation}}
\newcommand{\CR}{\notag \\}
\newcommand{\leqn}[1]{(\ref{#1})}

\def\Tr{\mathop{\rm Tr}}

\newcommand{\bspace}{\!\!\!\!}
\def\met{\mbox{$E{\bspace}/_{T}$}}

\def\stacksymbols #1#2#3#4{\def\theguybelow{#2}
    \def\vp{\lower#3pt}
    \def\sp{\baselineskip0pt\lineskip#4pt}
    \mathrel{\mathpalette\intermediary#1}}

\def\intermediary#1#2{\vp\vbox{\sp
     \everycr={}\tabskip0pt
     \halign{$\mathsurround0pt#1\hfil##\hfil$\crcr#2\crcr
              \theguybelow\crcr}}}

\def\gsim{\stacksymbols{>}{\sim}{2.5}{.2}}
\def\lsim{\stacksymbols{<}{\sim}{2.5}{.2}}
\begin{document}

\tikzset{
	  photon/.style={decorate, decoration={snake}, draw=black},
	  boson/.style={decorate, decoration={snake}, draw=black},
	  electron/.style={draw=black, postaction={decorate},
	           decoration={markings,mark=at position .55 with {\arrow[draw=black]{latex}}}
	  },
	  electron2/.style={draw=black, postaction={decorate},
	           decoration={markings,mark=at position .55 with {\arrow[draw=black]{latex reversed}}}
	  },
	  fermion/.style={draw=black, postaction={decorate},
	            decoration={markings,mark=at position .55 with {\arrow[draw=black]{}}}
	  },
 pttt/.style={decorate, draw=white},
	  gluon/.style={decorate, draw=black, %kred
	    decoration={coil,amplitude=4pt, segment length=6pt}},
	  gluon/.style={decorate, draw=black, %kred
	    decoration={coil,amplitude=4pt, segment length=6pt}},
	  higgs/.style={draw=black, postaction={decorate},
	           decoration={markings,mark=at position .55 with}
	  },
	  nothing/.style={draw=white}
	}

\title{Odd Top Partners at the LHC}

\author{Archana Anandakrishnan,}
\author{Jack H Collins,}
\author{Marco Farina,}
\author{Eric Kuflik,}
\author{and Maxim Perelstein}

\affiliation{Department of Physics, LEPP, Cornell University, Ithaca, NY 14853, USA}

\emailAdd{archana@physics.osu.edu}
\emailAdd{jhc296@cornell.edu}
\emailAdd{mf627@cornell.edu}
\emailAdd{kuflik@cornell.edu}
\emailAdd{mp325@cornell.edu}

\date{\today}

\abstract{LHC searches for fermionic top partners $T$ focus on three decay topologies: $T\to bW$, $T\to tZ$, and $T\to th$. However, top partners may carry new conserved quantum numbers that forbid these decays. The simplest possibility is a conserved parity, under which the top partner is odd and all SM states are even. In this case, decays of top partners may involve new particle-odd scalars, leading to signal topologies more commonly associated with supersymmetry, either with or without R-parity conservation. We study a simplified model in which this possibility is realized, and estimate the bounds on the top partner mass in this model implied by LHC searches for supersymmetry. We find that the bounds can be significantly weaker than in the conventional top partner decay scenario. For example, if the new parity is exact, a 500 GeV top partner is allowed as long as the lightest parity-odd scalar mass is between 325 and 500 GeV. The lower allowed top partner mass reduces the need for fine-tuning in the Higgs mass parameter, compared to the conventional decay scenario. We also present an explicit model, the Oddest Little Higgs, which exhibits this phenomenology.\\
}

%\keywords{}

%\arxivnumber{14xx.xxxx}

\preprint{}

\maketitle

\section{Introduction}

Many well-motivated extensions of the Standard Model (SM) at the weak scale contain ``top partners", particles that cancel the quadratic divergence in the top loop contribution to the Higgs mass parameter. Quantum numbers of the top partners are somewhat model-dependent. In a large class of SM extensions, including Little Higgs models~\cite{ArkaniHamed:2002qy} and five-dimensional ``Pseudo-Goldstone Higgs" models~\cite{Contino:2003ve} (see~\cite{Schmaltz:2005ky,Perelstein:2005ka,Bellazzini:2014yua,DeSimone:2012fs} for reviews), the top partner is {\it fermionic} (spin-1/2), {\it colored} (fundamental representation of the SM $SU(3)_C$), has an electric charge of $+2/3$, and is mostly an $SU(2)_W$ singlet. This particular species of top partner will be the focus of this paper.

Collider phenomenology of the top partner is largely determined by its mass and its quantum numbers. A fermionic top partner $T$ in the $({\bf 3}, {\bf 1})_{+2/3}$ representation of the SM gauge group is expected to be pair-produced at the LHC through QCD interactions, and decay to $t Z$, $t h$, and $b W$, with branching ratios of $25$\%, $25$\%, and $50$\%, respectively, fixed by the Goldstone boson equivalence theorem~\cite{Han:2003wu,Perelstein:2003wd}. LHC experiments have pursued dedicated searches for these processes, and their non-observation places a strong lower bound on the top partner mass: roughly, $m_T\gsim 800$ GeV from a recent ATLAS search based on 19.5 fb$^{-1}$ of 8 TeV data~\cite{ATLAS-CONF-2015-012} (see also~\cite{Chatrchyan:2013uxa} %~\cite{ATLAS:2013ima} from ATLAS
from CMS). These bounds, together with the discovery of a 125 GeV Higgs boson, rule out the most natural parameter region of the model. The required fine-tuning can be estimated as (see, {\it e.g.},~\cite{Berger:2012ec})
\beq
\Delta \approx \frac{3\lambda_t^2 m_T^2}{4\pi^2 m_h^2} \log \frac{\Lambda^2}{m_T^2} \gsim 10,
\eeq{FT}
where $m_T$ is the top partner mass, and $\Lambda\sim 10$ TeV is the cutoff scale of the model. It seems that top partners of this kind are increasingly endangered, at least if naturalness is to be taken seriously as a guide to the new physics landscape.

This conclusion may need to be modified, however, if top partners do not decay according to the pattern assumed in the LHC searches. This is the possibility that we investigate here. Deviating from the standard top partner decay pattern requires two ingredients. Firstly, $T$ needs alternate particles to decay to. Secondly, the couplings leading to the standard decays need to be suppressed. The first objective can be achieved by using global symmetry breaking patterns which contain more pseudo-Nambu-Goldstone bosons (pNGBs) than just the Higgs. This opens up the possibility that the top partner decays into a top and a neutral pNGB, or bottom and charged pNGB. The second objective can be achieved by implementing an approximate parity symmetry, under which all SM particles are even, and the top partner and the new pNGBs are odd.\footnote{Models with non-standard top partner decays have been previously considered, for example, in Refs.~\cite{Kearney:2013oia,Kearney:2013cca,Leskow:2014kga}; however, those models did not include a parity symmetry, so that both standard and non-standard $T$ decays were allowed. In contrast, here we will study models in which $T\to t Z$, $t h$, and $b W$ decays are forbidden by symmetry.} The possibility that the top partner which cancels the quadratic divergences coming from top loops is odd under such a parity was first considered in~\cite{Cheng:2005as}, in the context of Little Higgs models with T-parity~\cite{Cheng:2003ju,Cheng:2004yc}. In the case of an exact symmetry, heavy odd states will decay into light odd states and SM particles, and the lightest odd state would be stable. In the presence of small parity breaking, the lightest odd state will decay. We therefore consider Lagrangians with the generic form:
\beq
\mathcal{L} = \mathcal{L}_{\text{even}} + \epsilon \mathcal{L}_{\text{odd}}\,,
\eeq{equ:schematiclagrangian}
where $\mathcal{L}_\text{even}$ contains all of the parity preserving interactions, including all of the SM couplings and those required for the cancelation of quadratic divergences from top loops. $\mathcal{L}_{\text{odd}}$ contains all parity breaking interactions, which will be responsible for the decay of the lightest odd particle. The spurion $\epsilon$ schematically represents the size of the parity violation. Since there is an enhanced symmetry in the limit $\epsilon \to 0$, it is technically natural for these couplings to be very small.

In this paper we do not consider explicit extensions of the gauge sector which cancel the quadratic divergences coming from gauge boson loops. In the absence of new states associated with the gauge sector below a few TeV, there remains a residual little hierarchy problem in the gauge sector. This possibility was explored in~\cite{Katz:2005au}, if the cutoff is not low. Alternatively, the cancelation of divergences in the gauge sector can be decoupled from that in the fermion sector by having two symmetry breaking scales~\cite{Schmaltz:2010ac}, or by introducing supersymmetry at an intermediate scale (in which case the cancelation is achieved as in the MSSM by gauginos). In both of these cases, the new states can have masses of order a few TeV, without introducing significant fine tuning in the Higgs potential. Our choice of focusing only on the top and scalar sector is motivated by simplicity in the effective theory, but our models could derive from a UV completion in any of these categories.

\begin{figure}
\centering
\begin{tikzpicture}
\definecolor{oddline}{RGB}{200,110,30};
\definecolor{evenline}{RGB}{0,60,95};
\tikzstyle{evenlinestyle}=     [dash pattern=on 5pt off 3pt];

\draw [thick](-0.5,3) -- (-0.5,-1);
\node[anchor=east] (v1) at (-0.75,1) {1 TeV};
\draw [thick](v1) -- (-0.5,1);
\node[anchor=east] (v2) at (-0.75,2.5) {few TeV};
\draw [thick](v2) -- (-0.5,2.5);
\node[anchor=east] (v3) at (-0.75,-0.75) {200 GeV};
\draw [thick](v3) -- (-0.5,-0.75);

\node[anchor=west] (v4) at (1,0.5) {$T$};
\draw [thick,oddline](0,0.5) -- (v4);
\node[anchor=west] (v5) at (3.5,0) {$\omega, \eta$};
\draw [thick,oddline](v5) -- (2.5,0);
\draw [thick,oddline](3.5,0.1) -- (2.5,0.1);
\draw [thick,oddline](3.5,-0.1) -- (2.5,-0.1);

\node[anchor=west] (v6) at (3.5,0.75) {$\phi$};
\draw [thick,oddline](v6) -- (2.5,0.75);
\node[anchor=west] (v7) at (1,2.5) {$T',Q'$};
\draw [thick,oddline](0,2.5) -- (v7);
\draw [thick,evenlinestyle,evenline](0,2.65) -- (1,2.65);
\draw [thick,evenlinestyle,evenline](0,2.35) -- (1,2.35);
\node[anchor=west] (v8) at (3.5,2.5) {$W',Z'$};
\draw [thick,evenlinestyle,evenline](2.5,2.5) -- (v8);
\draw [thick,evenlinestyle,evenline](2.5,2.65) -- (3.5,2.65);
\draw [thick,oddline](2.5,2.35) -- (3.5,2.35);
\end{tikzpicture}
\caption{The type of spectrum considered in this paper. Solid, orange lines represent parity-odd particles, while parity-even states are represented by dashed blue lines. We assume that LHC phenomenology is dominated by a set of parity-odd states below the TeV scale, including a single top partner responsible for the cancelation of quadratic divergences, and a set of scalars that allow it to have interesting phenomenology. There may be additional fermionic and bosonic states at a multi-TeV scale associated with a UV completion of this picture.}
\label{fig:generalspectrum}
\end{figure}
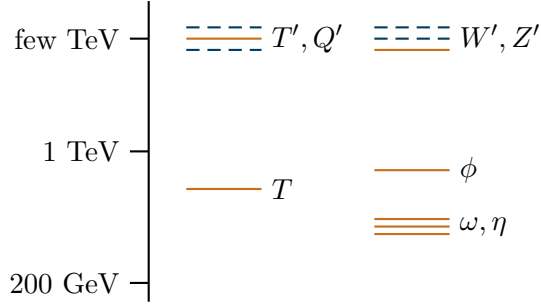

The rest of the paper is organized as follows. A simplified model that encompasses the LHC phenomenology of interest to us, and the three particular top partner decay scenarios that occur naturally in this model, are presented in Section~\ref{sec:simpmodels}. The LHC constraints on this simplified model, within each of the $T$ decay scenarios, are studied in detail in Section~\ref{sec:recasts}. In Section~\ref{sec:OddestModel}, we describe the ``Oddest Little Higgs", a complete non-linear sigma model (NLSM) that gives rise to the simplified model, and hence the LHC phenomenology, considered in the first half of the paper. We summarize our conclusions in Section~\ref{sec:conc}.

\section{Simplified Models}
\label{sec:simpmodels}

We suppose that the Higgs is a pNGB of a spontaneously broken approximate global symmetry, and extend the SM top sector so that the top Yukawa couplings only break the global symmetry collectively, eliminating the one-loop quadratic divergence in the Higgs mass  parameter. We further assume that the non-linear sigma model (NLSM) that encodes the global symmetry breaking, as well as the extended top sector, are invariant under a parity such that all SM particles are even and any new particles with masses $\lesssim 1 \; \text{TeV}$ are odd. We refer to this extra symmetry as ``t-parity", to distinguish it from the conventional T-parity; our effective theory does not contain any new states in the gauge sector, allowing for simpler implementation of parity compared to the conventional LHT models. In Section~\ref{sec:OddestModel}, we will present an explicit theory, the ``Oddest Little Higgs" (OLH), that satisfies these requirements and is phenomenologically viable. First, however, we would like to focus on the LHC signatures of this class of models, using a simplified model approach.

Below the TeV scale, our model contains a vector-like pair of fermionic top partners, $T$ and $T^c$; and additional scalars which are pNGBs of the global symmetry, $\eta$ and $\omega$. Their quantum numbers of these states under the SM SU(2)$_L$ $\times$ U(1)$_Y$ gauge symmetry are as follows:
\begin{equation}
T, T^c: ~{\bf 1}_{\pm 2/3},~~~~~~~~ \eta: ~ {\bf 1}_0,~~~~~~~~ \omega: ~ {\bf 3}_0.
\end{equation}
We assume that one of the electrically neutral scalars, either $\eta$ or $\omega^0$, is the lightest t-odd particle (LtP); otherwise, strong LHC limits on stable charged particles~\cite{Chatrchyan:2013oca} would apply if the LtP were both charged and long lived.

The LHC phenomenology is described by the following simplified Lagrangian:
\beqa
\mathcal{L} &=& \mathcal{L}_\text{SM} + \mathcal{L}_\text{Kin} + \frac{1}{2}m_\eta^2 + m_\omega^2 \text{Tr}\left[\omega^2\right] + m_T T T^c + y_\eta T t^c \eta +  \frac{y_\omega}{f} (Q_L^3 \omega H) T^c\label{equ:simplag}\CR
& &~~~~~~~~~+ \frac{1}{f} \left(Q_L \boldsymbol{\epsilon}_\eta^{(u)} {u^c} H \eta + Q_L \boldsymbol{\epsilon}_\eta^{(d)} {d^c} H^* \eta   + Q_L \boldsymbol{\epsilon}_\omega^{(u)} {u^c}\omega H + Q_L \boldsymbol{\epsilon}_\omega^{(d)} {d^c}\omega H^* \right).%\notag
\eeqa{eq:SimpMod}
Here, $f$ is the mass scale at which the non-renormalizable interactions of the model are generated; in the OLH model, it is identified with the ``pion decay constant" of the NLSM.
The t-parity preserving couplings in the first line arise in the OLH model from the same operators responsible for the top Yukawa, and generically $y_\eta$, $y_\omega\sim\mathcal{O}(1)$. We assume that the similar parity-preserving couplings involving the light quarks are Yukawa suppressed and negligible. The couplings in the second line of Eq.~(\ref{equ:simplag}) encode the possibility of small t-parity violation; in the presence of these couplings, the LtP can decay to SM quarks, leading to interesting LHC signatures. The  $\boldsymbol{\epsilon}$ couplings are matrices in flavor space and are not related to the SM Yukawas, and therefore have much more freedom in their flavor structure. The most flavor-safe structures would be minimally flavor-violating (MFV) or universal, but anarchic and inverted structures are also possible so long as the overall scale of these spurions is sufficiently small to avoid flavor constraints. This is technically natural due to the enhanced symmetry when all of these couplings are set to zero. The LHC constraints will generally be weakest when the decay products are light jets, and for simplicity we will assume that the LtP either decays exclusively to first generation quarks, or is stable on detector time scales and neutral.

At the LHC, the t-odd top partners will be pair produced with a QCD production cross section~\cite{Aliev:2010zk}. Unlike the traditional top partners, the single production of such partners is forbidden by t-parity. (T-violating interactions may induce single production cross section of order $\epsilon^2$; we assume that this is too small to play a role in the LHC phenomenology.) The experimental signatures of the t-odd top partner are model-dependent, since a variety of decay patterns are possible. Three phenomenologically distinct, simple scenarios can be realized by the Lagrangian of Eq. (\ref{equ:simplag}):

\begin{figure}
\centering
\resizebox {\textwidth} {!} {
\begin{tikzpicture}

\node[anchor=west] at (-4,4) {\textbf{Scenario 1}};
\node[anchor=east] (v1) at (-4,3) {$m_T~$};
\draw [thick] (v1) edge (-3,3);
\node[anchor=east] (v3) at (-4,0) {$m_T-m_t~$};
\draw [dashed, thick] (v3) edge (-1.5,0);
\node[anchor=west] (v6) at (-1.5,-1.5) {$~ m_\eta$};
\draw [thick]  (-2.5,-1.5) edge (v6);
\draw [thick] plot[smooth, tension=.7] coordinates {(-3.5,3)};
\draw [thick](2.5,3);
\draw [thick,-triangle 45](-3.5,3) .. controls (-3.5,1) and (-2,2) .. (-2,-1.5);
\node[anchor=west] at (-2.5,1) {$T \to t \eta$};
\draw [thick](-2.5,2.6) -- (-1.5,2.6)node[anchor=west] {$m_\omega$};
\draw [thick](-2.5,2.7) -- (-1.5,2.7);

\begin{scope}[shift={(6,0)}]
\node[anchor=west] at (-4,4) {\textbf{Scenario 2}};
\node[anchor=east] (w1) at (-4,3) {$m_T~$};
\node (w2) at (-3,3) {};
\draw [thick] (w1) edge (-3,3);
\node[anchor=east] (w3) at (-4,0) {$m_T-m_t~$};
\node (w4) at (-3,0) {};
\draw [dashed, thick] (w3) edge (-1.5,0);
\draw [thick](-2.5,3) -- (-1.5,3)node[anchor=west] {$m_\eta$};
\node[anchor=east] at (-2.75,1.75) {$T \to b \omega^+~$};

\node[anchor=west] (w6) at (-1.5,1) {$~ m_{\omega^+}$};
\draw [thick] (-2.5,1) edge (w6);
\draw [thick] plot[smooth, tension=0.7] coordinates {(-3.5,3)};
\draw [thick](-3.5,3);
\draw [thick,-triangle 45](-3.5,3) .. controls (-3.5,1.5) and (-2,2.5) .. (-2,1);
\end{scope}

\begin{scope}[shift={(11.5,0)}]
\node[anchor=west] at (-4,4) {\textbf{Cascade}};
\node[anchor=east] (w1) at (-4,3) {$m_T~$};
\node (w2) at (-3,3) {};
\draw [thick] (w1) edge (-3,3);
\node[anchor=east] (w3) at (-4,0) {$m_T-m_t~$};
\draw [dashed, thick] (w3) edge (-1.5,0);
\draw [thick](-2.5,0.5) -- (-1.5,0.5)node[anchor=west] {$m_\eta$};
\node[anchor=east] at (-3,2) {$T \to b \omega^+$};
\node[anchor=east] at (-2,1) {$\omega^+ \to \eta \bar{q} q~$};

\node[anchor=west] (w6) at (-1.5,1.5) {$~ m_{\omega^+}$};
\draw [thick] (-2.5,1.5) edge (w6);
\draw [thick] plot[smooth, tension=0.7] coordinates {(-3.5,3)};
\draw [thick](-3.5,3);
\draw [thick,-triangle 45](-3.5,3) .. controls (-3.5,2) and (-2,2.5) .. (-2,1.5) {};
\draw [thick, -triangle 45] (-2,1.5) -- (-2,0.5);
\end{scope}

\end{tikzpicture}
}
\caption{Decay scenarios depending on the mass hierarchies. The decay $T \to t \eta$ will typically dominate if it is kinematically allowed (scenario 1). If $m_\eta > m_T - m_t$, then the decay $T \to  b\omega^+$ will dominate if it is allowed (scenario 2).  If $m_T > m_\omega > m_\eta > m_T - m_t$, then cascade decays may be typical.}
\end{figure}
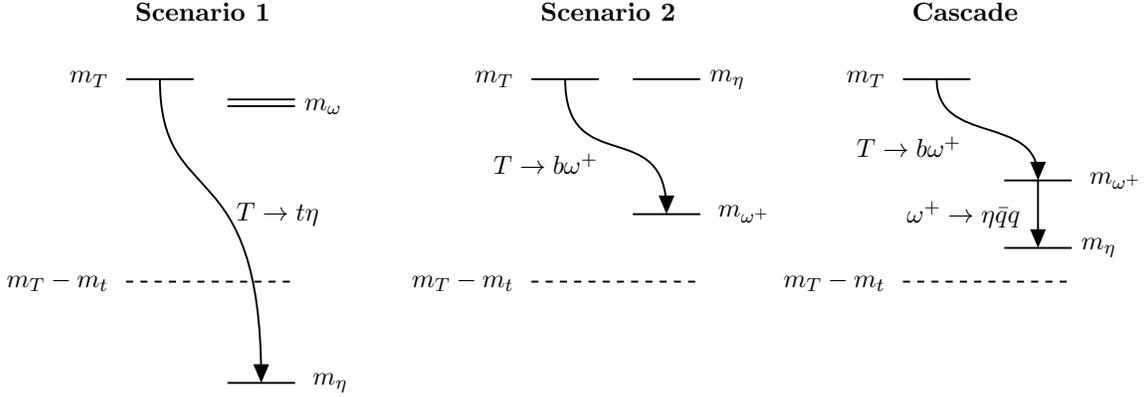

\subsubsection*{Scenario 1: Singlet LtP}

In the OLH model, it is natural for $\eta$ to be the LtP, since $\omega$ receives quadratically divergent contributions to its mass from gauge loops, while $\eta$ does not. If this is the case, the decay $T \to t \eta$ will typically dominate. (Even if decays to $\omega$ are kinematically accessible, the corresponding couplings are suppressed by a factor of $\mathcal(v/f)$.)  If t-parity is exact, $\eta$ is a stable, weakly-interacting particle, leading to a SUSY-like signature $t\bar{t}+\met$, see Section~\ref{sec:met}. If t-parity is approximate so that the decay $\eta\to jj$ is allowed, the final state is instead $t \bar{t} j j j j$, with two jet pairs forming resonances with the same mass, $m_\eta$. The $\eta$ decays may be either prompt or displaced, depending on the value of $\epsilon$. Hadronic decays of the top can result in final states with 10 hard jets (including two $b$'s), potentially more with additional hard gluon emissions. This scenario will therefore be strongly constrained by multi-jet R-Parity Violating (RPV) gluino searches, as we discuss in Section~\ref{sec:topsjets}.

\subsubsection*{Scenario 2: Triplet LtP}

Since the size of the UV contributions to the scalar masses is not calculable, we should also consider the possibility that $\omega$ is the LtP. In this case, if $T\to t\eta$ is not kinematically available, the top partner will decay via $T \to t \omega^0$ or $T \to b \omega^+$. The first of these decays leads to the same phenomenology as scenario 1. However, if $m_T-m_b > m_\omega > m_T - m_t$, the decay $T \to b \omega^+$ dominates. Radiative corrections and non-renormalizable operators in the OLH model inevitably induce a small splitting, typically $\mathcal{O}$(10 MeV), between the $\omega$ states. We assume that $\omega^0$ is the LtP, in which case $\omega^\pm$ will decay to $q\bar{q}^\prime \omega^0$ or $\ell^\pm \nu \omega^0$; however, the jets and leptons produced in these decays are too soft to be detected. If t-parity is exact, this scenario results in a signature $b\bar{b}+\met$, covered by SUSY searches, see Section~\ref{sec:bMET}. If t-parity is approximate, the $b\bar{b}jjjj$ final state is produced, and constraints from multi-jet searches will apply. However if the $T$-$\omega$ mass splitting is small, the $b$ jets will typically be soft, relaxing the constraints from such searches. This will be discussed in Section~\ref{sec:compressed}.

\subsubsection*{Scenario 3: Cascade Decays}

Finally, it is also possible that both $\eta$ and $\omega$ are light enough to participate in the decays of the top partner, leading to cascade decays and complex, high-multiplicity signatures. For example, the chain $T\to b\omega^+$, $\omega^+\to q\bar{q}^\prime\eta$, may  produce a $b\bar{b}+4j+\met$ final state, if the t-parity is exact, or a $b\bar{b}+8j$ final state, if it is approximate. Some of the jets may be soft depending on the $T$-$\omega$ and $\omega$-$\eta$ mass splittings.

\subsection{Electroweak Precision Constraints on the Simplified Model}

Electroweak precision data place significant constraints on the parameter space of models with fermionic top partners, which need to be taken into account in any discussion of direct searches. For example, in Littlest Higgs models, based on the same coset as our OLH model, potentially large tree-level contributions to electroweak precision observables arise from the vacuum expectation value (vev) of the triplet scalar, and from $Z^\prime$ exchange diagrams~\cite{Csaki:2002qg}. Neither of these effects is present in our model: triplet vev is forbidden by t-parity, while $Z^\prime$ bosons do not appear at the scale $f$. Moreover, the leading one-loop contributions to the electroweak precision observables that dominate the constraints in the Littlest Higgs model with T-parity~\cite{Hubisz:2005tx} are also absent, since those loops involve parity-even top partners absent in our model~\cite{Cheng:2005as}. Thus, we expect the precision electroweak constraints on our model to be quite weak.

Here, we consider the contributions to precision electroweak observables produced by the particles and interactions of the simplified model, Eq.~\leqn{eq:SimpMod}. These are in a sense ``irreducible", since they follow directly from the structure that gives rise to the LHC signatures of interest to us. It turns out that these contributions are in fact quite small, allowing the t-odd top partners to be as light as $300$ GeV. Of course, a more complete description of the physics that gives rise to Eq.~\leqn{eq:SimpMod} will generally introduce additional, model-dependent contributions to precision electroweak observables; we leave an analysis of those contributions in the OLH model for future work. 

Starting with Eq.~\leqn{eq:SimpMod} and integrating out the heavy top partner and $\omega$ triplet leads to one-loop corrections to the top $Z \bar{b}_L b_L$ vertex. Following the conventions of~\cite{Bamert:1996px}, the corrections to coupling are
\beq
\delta g_L^b \simeq \frac{|y_\omega|^2}{32\pi^2} \frac{v^2}{f^2} \left[\left(-\frac{1}{2} +s_w^2 \right)\log \frac{\Lambda^2}{m_T^2} +  \left(-\frac{1}{2} + \frac{4}{3}s_w^2 \right)\log \frac{\Lambda^2}{m_\omega^2} \right] + \rm finite.
\eeq
where the coupling $\delta g_L^b$ is defined by effective Lagrangian $\mathcal{L}_{\rm eff} = \frac{e}{s_w c_w} Z_\mu (g_L^b + \delta g_L^b  ) \bar{b}_L \gamma^\mu b_L$ and $g_L^b=\frac{1}{2}+\frac{1}{3}s_w^2$ is the SM coupling. The divergence indicates that there is a counterterm somewhere in the full theory, that can contribute to $\delta g_L^b$ but is incalculable within the chiral Lagrangian. We can still get an estimate on the constraint, requiring that the above contribution not be too large for $\Lambda = 4\pi f \sim 4\sqrt{2}\pi m_T$, where in the last step we used the relation $m_T \approx f/\sqrt{2}$ obtained in the OLH model.

The SM prediction from electroweak precision fits and the measurements from LEP~\cite{ALEPH:2005ab} are
\[
g_L^b ({\rm SM}) = -0.42114^{+0.00045}_{-0.00024},~~~~~~~~
g_L^b ({\rm LEP}) = -0.4182^{+0.0015}_{-0.0015}\, .
\]
The one-loop contribution can only worsen the fit. Requiring that the top partner does not contribute another $2\sigma$ deviation from the SM prediction constrains
\beq
y_{\omega} \frac{v}{f} \lesssim 0.58\,.
\end{equation}
Given that generically $y_\omega\sim 1$, this bound is satisfied for $f\gsim 500$ GeV, or (again using $m_T \approx f/\sqrt{2}$) for $m_T\gsim 300$ GeV.

The light scalar triplet, $\omega$ can contribute logarithmically divergent contributions to the $W$ boson mass, if the masses of charged and neutral component are split. The corresponding contribution to the Peskin-Takeuchi oblique $T$ parameter~\cite{PhysRevLett.65.964,PhysRevD.46.381} is
\beq
\delta T = \frac{1}{2\pi s_w^2 c_w^2}\frac{\delta m^2_\omega}{m_Z^2} \log \frac{\Lambda^2}{m_{\omega_+}^2},
\end{equation}
where $\delta m^2_\omega\equiv m_{\omega^+}^2-m_{\omega^0}^2$. The current bounds on $T$ constrain $\delta m^2_\omega \lesssim 200~\rm{GeV}^2$. A general UV-completion can be expected to generate mass-splitting $\delta m^2_\omega \sim a \times \frac{v^4}{f^4} m^2_{\omega_0} $, where $a$ is a model-dependent numerical factor. In the OLH model presented in Section~\ref{sec:OddestModel}, we find $a = \frac{1}{16}$ at leading order. Assuming $m_{\omega^0}\sim v$, as will be typical for the phenomenological scenarios considered here, we find a bound $f \gsim 500$ GeV, corresponding to $m_T\gsim 300$ GeV in the OLH model.

\section{Bounds from 8 TeV LHC}
\label{sec:recasts}

In this section, we estimate the current LHC bounds on the different topologies described above. To do so we recast searches performed by both ATLAS and CMS, mainly in the context of supersymmetric models. In all cases the $ p p \rightarrow T \bar{T}$ process has been simulated with
{\tt  MadGraph5 aMC@NLO 2.2.3}~\cite{madgraph}, using CTEQ6L parton distribution functions~\cite{Pumplin:2002vw}, followed by decaying, showering and detector simulation performed through {\tt Pythia 6.4}~\cite{pythia1,pythia2} and {\tt PGS4}~\cite{PGS}. After all cuts the LO cross sections have been rescaled by a K-factor extracted from \cite{Aliev:2010zk}, which amounts to a factor $\sim 1.5$ in most of the mass range under consideration.

The following sections describe in detail the recast searches. Each is characterized by a fixed decay channel for the top partner, either $T \rightarrow t \eta$ or $T \rightarrow b \omega^+$, and by the properties of the scalar involved in the decay chain, in particular whether it is stable or decays promptly. We do not consider the case where LtP lifetime corresponds to displaced decays inside a detector, since displaced decays into jet pairs are very strongly constrained at the LHC independent of the details of the event~\cite{Aad:2013ija}. In all scenarios we assume $100\%$ branching ratio in the channels of interest for both $T$ and the scalars.

\subsection{Scenario 1: $T \bar{T} \to t \bar{t} \eta \eta $}
\label{sec:recast1}

If the singlet $\eta$ is the LtP, the decay $T\to t\eta$ dominates. We consider two cases: exact t-parity (stable LtP) and broken t-parity (unstable LtP).

\begin{figure}
\begin{center}
  \includegraphics[width=0.45\textwidth]{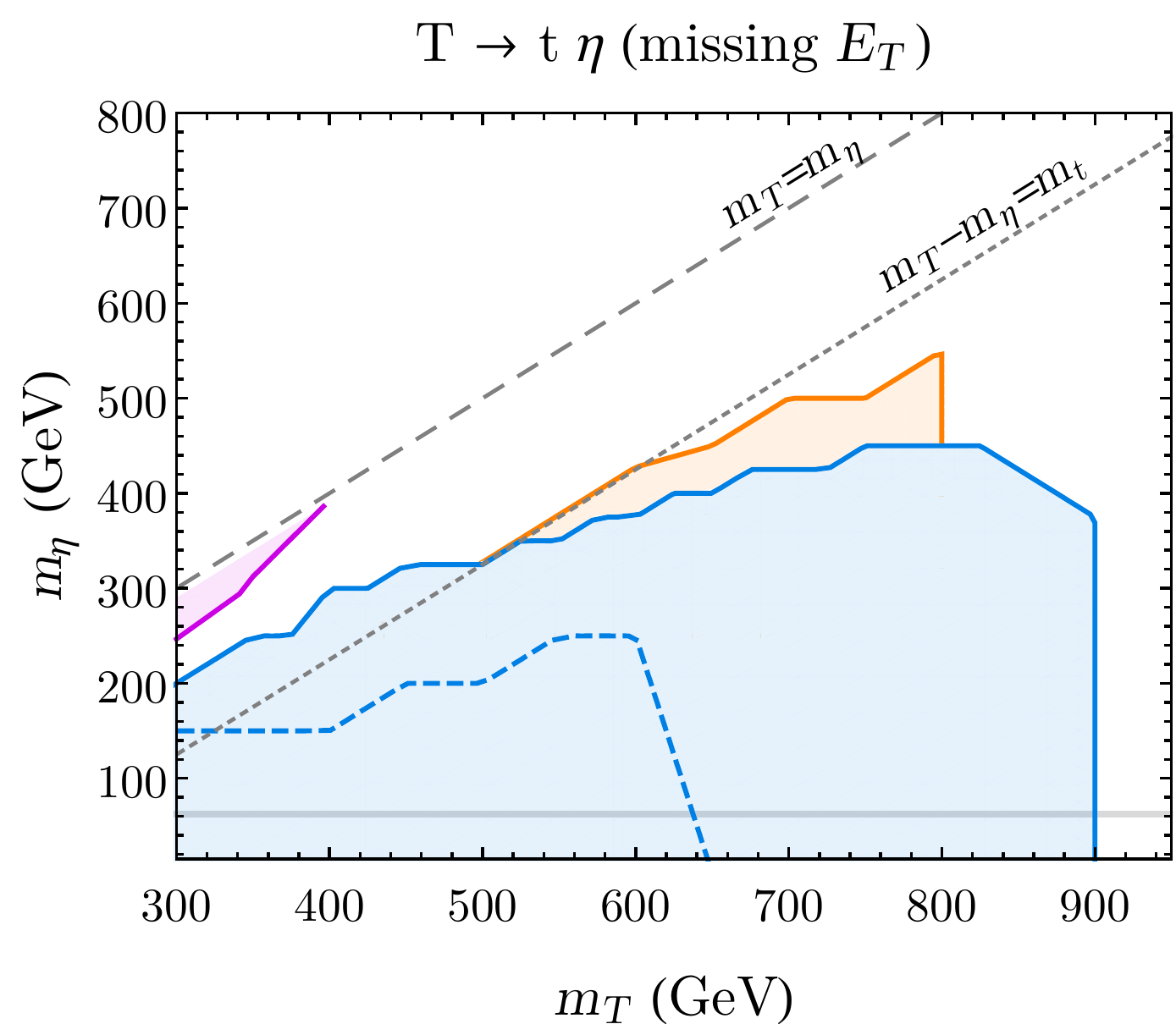} \hspace{1cm}
  \includegraphics[width=0.45\textwidth]{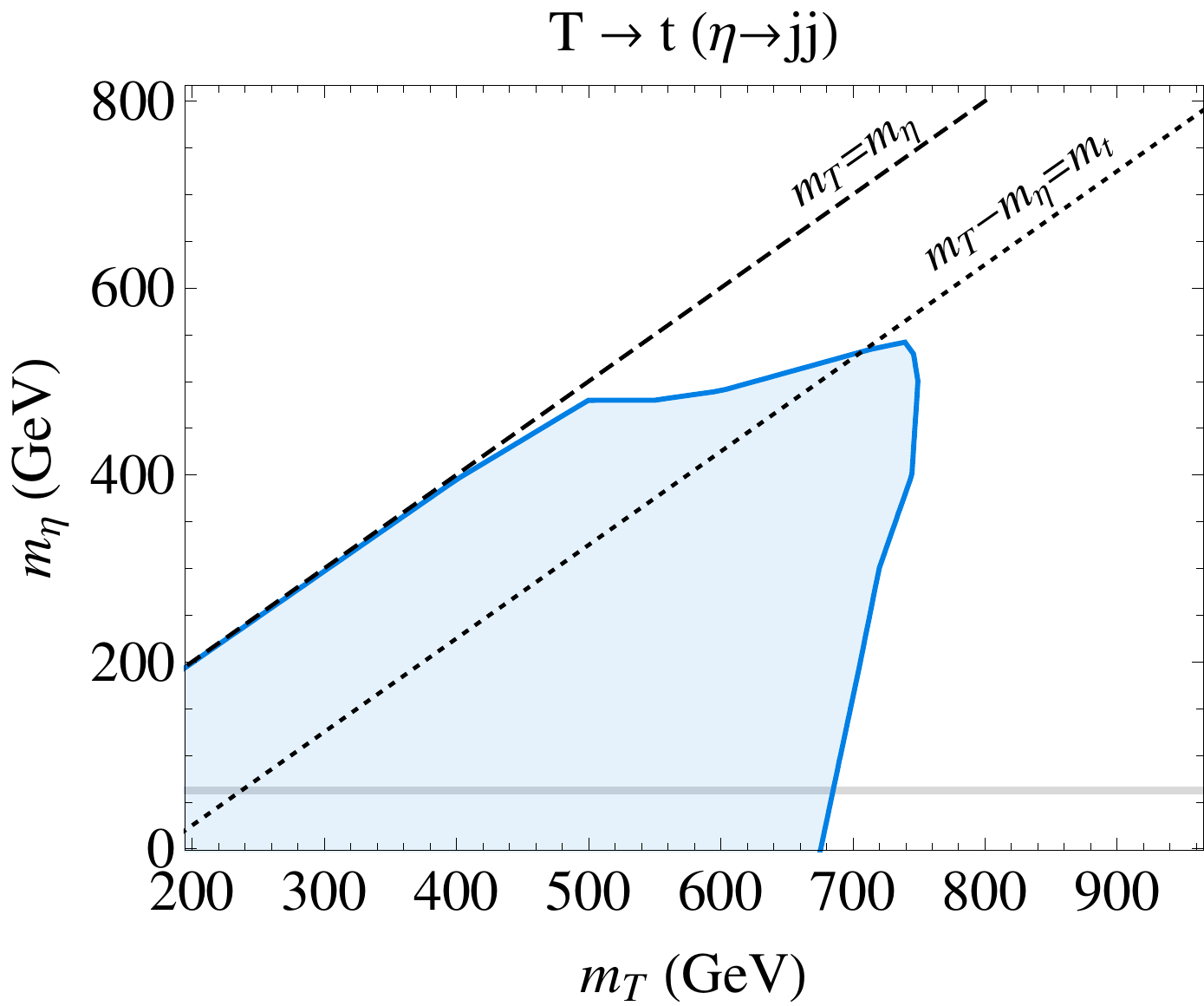}
  \end{center}
 \caption{LHC bounds for scenario 1, $T \to t \eta$. Left panel: Exact t-parity case. The blue/orange shaded areas are excluded by the CMS~\cite{Khachatryan:2015pwa}/ATLAS~\cite{Aad:2014kra} searches for isolated lepton, jets, and missing transverse momentum, assuming the same acceptance and cut efficiency for spin-1/2 and spin-0 signal models. The dashed line indicates the bound from the CMS cut-and-count search in the same channel~\cite{Chatrchyan:2013xna}, including the difference in the cut efficiencies. The purple area is excluded by the mono-jet search \cite{Aad:2014nra}. Right panel: Approximate t-parity case, $\eta\to jj$. The blue shaded area is excluded by the ATLAS multijet analysis~\cite{Aad:2015lea}. In both panels, below the horizontal gray line the Higgs decay $h \rightarrow \eta \eta$ is kinematically accessible. \label{fig:recast1}}
\end{figure}

\subsubsection{Exact t-Parity}
\label{sec:met}

The signal topology in this case is identical to that of stop squark ($\tilde{t}$) pair-production, where the stop decays via $\tilde{t} \to t \tilde{N}$ and $\tilde{N}$ is a stable neutralino. Many searches for this SUSY process have been performed at the LHC. In the region of the parameter space where a two-body decay to $t\tilde{N}$ is kinematically allowed, the strongest bounds can be derived from the ATLAS and CMS searches for isolated lepton, jets, and missing transverse momentum (MET)~\cite{Aad:2014kra,Khachatryan:2015pwa,Chatrchyan:2013xna}. The ATLAS collaboration supplies acceptances and efficiencies to pass the selection cuts as a function of $m_{\tilde{t}}$ and $m_{ \tilde{N}}$ for $m_{\tilde{t}}<800 \; \text{GeV}$. We assume that these acceptances and efficiencies apply to the fermionic top partners as well, with $m_T=m_{\tilde{t}}$ and $m_\eta=m_{ \tilde{N}}$. This assumption ignores the differences in the kinematic distributions of the fermionic and scalar top partners; we will comment on this effect below. We then use the calculated $T$ pair-production cross section and the 95\% C.L. bounds reported by ATLAS to place constraints on the $m_{T}$-$m_{\eta}$ plane, shown in the left panel of Fig.~\ref{fig:recast1} (solid orange line). Likewise, the CMS collaboration provides a 95\%~C.L. upper bound on the $pp\to\tilde{t}\tilde{t}^*$ cross-section, in $m_{\tilde{t}}-m_{ \tilde{N}}$ plane, for $m_{\tilde{t}}< 900$~GeV and $m_{\tilde{t}}-m_{\tilde{N}}> 100$~GeV. Neglecting the differences in kinematic distributions, we use the calculated $T$ pair-production cross section to obtain the bound shown in Fig.~\ref{fig:recast1} (solid blue line).
 
To test the effect of the differences in kinematic distributions of spin-1/2 top partner and stop signals, we compared the efficiency of the cut-and-count search presented in Ref.~\cite{Chatrchyan:2013xna} for the cases of the $T\to t\eta$ and ${\tilde t}\to t\tilde{N}$ signal models, for a grid of points in the parameter space. We find that across the parameter space, the efficiency is significantly {\it lower} in the case of the $T\to t\eta$ signal, compared to the ${\tilde t}\to t\tilde{N}$ signal with the same mother and daughter masses. The reason is that the spin-1/2 top partners on average have smaller production-frame velocity compared to stops of the same mass, due to a steeper rise of the cross section at the kinematic threshold in the spin-1/2 case. This translates into lower MET and lower $p_T$ of the visible decay products. The bound from the cut-and-count search~\cite{Chatrchyan:2013xna} on our model, including the effect of kinematic distributions, is shown by the dashed blue line in Fig.~\ref{fig:recast1}. Unfortunately, we were not able to evaluate the effect of kinematic distributions on the other relevant searches in this channel, since they involve advanced multivariate statistical techniques such as boosted decision trees. However, we note that for the case of stops, the bounds imposed by the cut-and-count search~\cite{Chatrchyan:2013xna} are only slightly weaker than those from the more complex searches. We expect the same to be true for the spin-1/2 top partner, meaning that the true bound is somewhat, but not dramatically, stronger than indicated by the dashed line. In any case, this analysis strongly suggests that the solid blue and orange lines in Fig.~\ref{fig:recast1} represent a very conservative interpretation of the data, and the true bounds are likely to be significantly weaker. We conclude that for a light LtP these searches can probe fermionic top partners up to $650 \; \text {GeV} \lesssim m_T \lesssim 1 \; \text{TeV}$, but in the compressed region with $m_T - m_\eta < 175 \; \text {GeV}$ their sensitivity is substantially degraded, leaving a window that is unconstrained. 

In this compressed region, constraints from the mono-jet search \cite{Aad:2014nra} become important. In this case, we use {\tt CheckMate} \cite{Drees:2013wra}, based on the fast detector simulation {\tt DELPHES 3} \cite{deFavereau:2013fsa} to recast the bounds in terms of our model. This procedure automatically takes into account the differences in kinematic distributions between our model and the case of stops. The excluded region is also shown in the left panel of Fig.~\ref{fig:recast1} (purple line). This search rules out very degenerate spectra below $m_T\approx 400$ GeV (which compares to the reach of $\approx 300\; \text{GeV}$ for stops), and does not impose any constraint for heavier top partners. The CMS search for soft leptons in association with initial-state radiation (ISR) jet and MET \cite{CMS-PAS-SUS-14-021}, may also be relevant in the compressed region. This search has a similar reach for stops as the ATLAS monojet search, and we expect the same is true for fermionic top partners. The compressed region is also probed by the ATLAS search in the $W^+W^-$ topology \cite{Aad:2015pfx} This analysis is sensitive for stops only in the region $m_{\tilde{t}}\lsim 200$ GeV, and while the top partner bound is probably somewhat stronger due to higher production cross section, the rapid decrease of the cross section with mass implies that this search does not constrain the masses of interest to us. Therefore we do not explicitly recast it in this work. We conclude that top partners with mass $m_T \gtrsim 400 \; \text{GeV}$ are not yet constrained by searches in this compressed regime, which compares to $\sim 300 \; \text{GeV}$ for stops.

\subsubsection{Approximate t-Parity}
\label{sec:topsjets}

The decay chain of interest in this case is $T \rightarrow t (\eta \rightarrow j j )$. Most searches with tops in the final states rely on the presence of extra leptons, as in the case of standard fermionic top partners decays in $tZ$ or $bW$, or rely on same-sign dileptons as typical in supersymmetric models involving stops. As such they do not apply to our case. We thus require the tops to decay hadronically and we recast an ATLAS analysis for massive particles decaying to multiple jets, designed to search for RPV gluinos~\cite{Aad:2015lea}.

The analysis requires $\geq6$ or $\geq7$ jets each with high $p_T$ and $|\eta|<2.8$. Different search regions are categorized by different $p_T$ cut and number of minimum required b-tagged jets.
In particular our signal at parton level is comprised of 2 b's and 8 jets. Given the presence of b's and the fact that intermediate state particles are on-shell, we find the most constraining search category to be the one requiring a minimum of 2 b-tags and $p_T > 80$ GeV for all $\geq7$ jets. The expected background is $1670 \pm190$ events, while $1560$ events have been observed during data taking, corresponding to $20.3$ fb$^{-1}$ of collected luminosity at 8 TeV.

First we compute the expected number of signal events for each point in parameter space. The signal likelihood is then estimated through the standard $CL_s$ technique, where we fix the expected background to its central value. The $95\%$ C.L. excluded area is shown in Figure~\ref{fig:recast1}, right panel. The upper bound on the top partner mass is at most $m_T \gtrsim 850$ GeV, and degrades to approximately 700 GeV in the light-LtP region $m_\eta =0$, and to as low as 500 GeV in the quasi-degenerate region $m_\eta \approx m_T$. In the former region, $\eta$ is produced with a large boost, so that the two jets stemming from its decay are often merged. This effect reduces the total number of jets of the final state, making it less likely to pass the $\geq7$ jets cut. In the latter region, the tops and the $\eta$'s are produced almost at rest in the lab frame, and thus their decays produce softer jets which often fail to pass the  $p_T > 80$ GeV cut.

Let us briefly comment on possible constraints in similar scenarios with different $\eta$ decays, namely into third generation quarks. If $\eta \rightarrow b \bar{b}$ we can expect the bounds to be somewhat stronger than in the light generation case, since the higher number of b's in the final state increases the probability of passing the b-tag cut, while the kinematics is nearly identical. If $\eta \rightarrow t \bar{t}$, an interesting six tops final state appears which is not directly addressed by any search at present. However, a recent recast~\cite{Deandrea:2014raa} points to bounds of the order $m_T \gtrsim 700$ GeV for most $\eta$ masses.

If the $\eta \rightarrow j j$ decay is long lived on detector scales, much stronger constraints coming from the CMS displaced dijet search \cite{Aad:2013ija} apply for lifetimes between 1 mm and 1 km. For the case where $m_\eta > m_T$, the topology is very similar to the displaced gaugino decay, $\tilde{g}/\tilde{N} \to jjj$, studied in \cite{Cui:2014twa,Schwaller:2015gea,Liu:2015bma,Csaki:2015uza}.

\subsection{Scenario 2: $T \bar{T} \to b \bar{b} \omega^+ \omega^-$}

\begin{figure}
\begin{center}
 \includegraphics[width=0.45\textwidth]{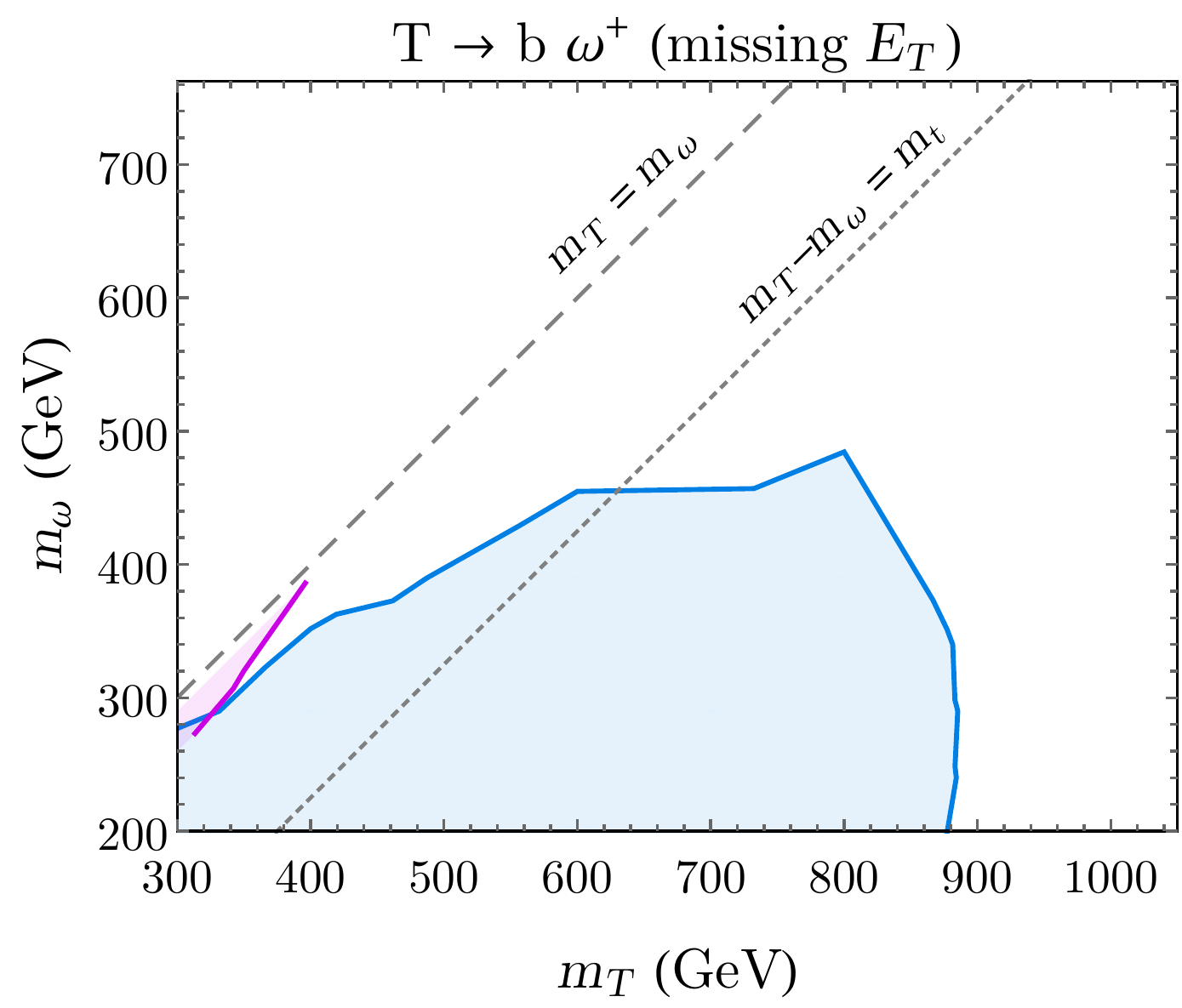}  \hspace{1cm}
 \includegraphics[width=0.45\textwidth]{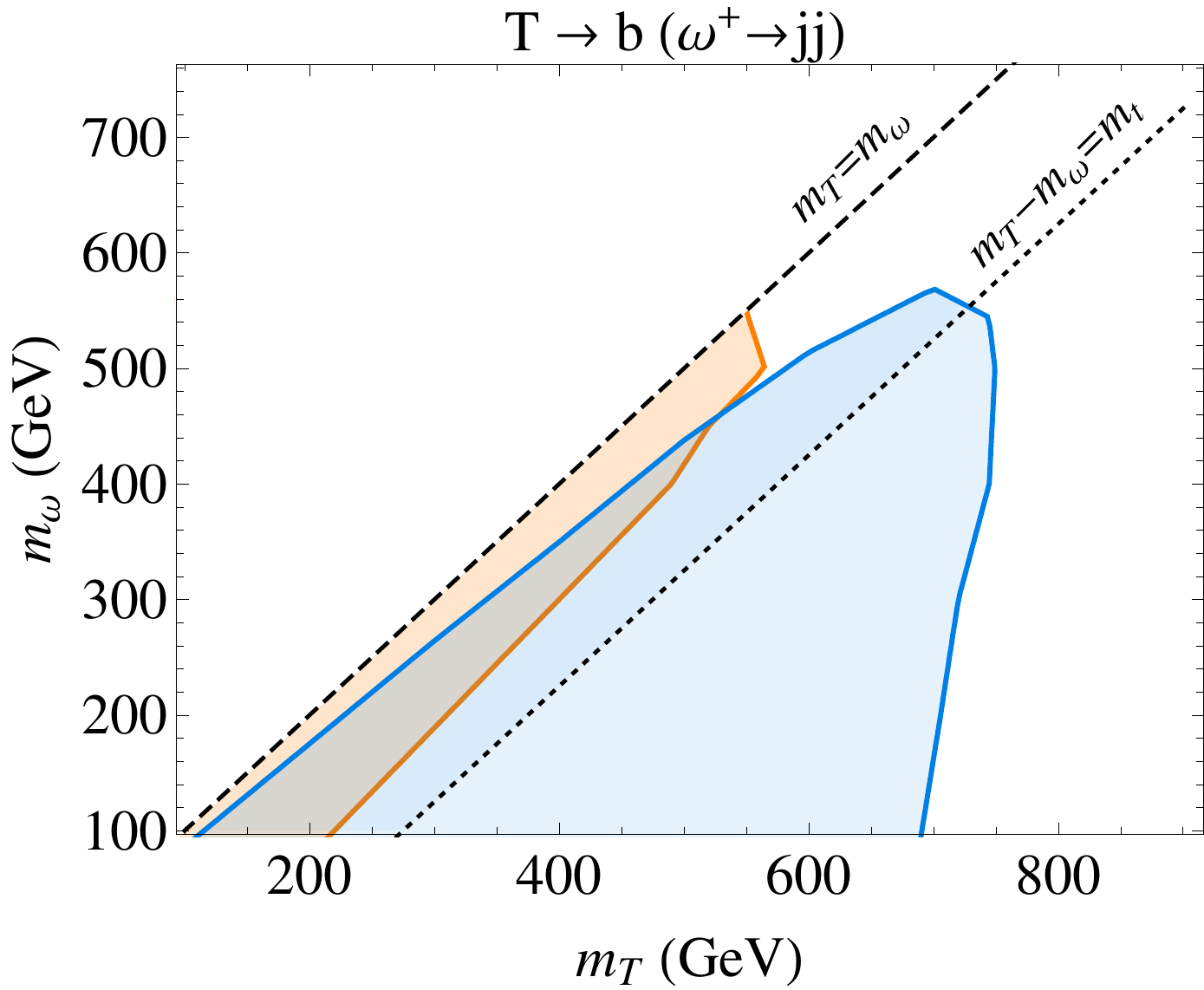}
  \end{center}
 \caption{LHC bounds for Scenario 2, $T \to b \omega^+$. Left panel: Exact t-parity case. The blue shaded area is excluded by the ATLAS search for 2 b-jets and $\met$~\cite{Aad:2013ija}. The purple area is excluded by the mono-jet search~\cite{Aad:2014nra}. Right panel: Approximate t-parity case, $\eta\to jj$. The blue shaded area is excluded by the ATLAS multijet analysis~\cite{Aad:2015lea}, while the red shaded area is excluded by the CMS dijet resonances search~\cite{Khachatryan:2014lpa}.
 \label{fig:recast2}}
\end{figure}

We next consider the scenario where $\omega^0$ is the LtP, and $\omega^+$ and $\omega^0$ nearly degenerate. In the case of exact t-parity, the $\omega^+$ decays to $\omega^0$ and soft leptons or jets, which are too soft to be detected. In the case of approximate t-parity, the direct decay $\omega^+ \to q \bar{q}$ is permitted along with the decay via an intermediate $\omega^0$. Both of these channels are phenomenologically equivalent, appearing as $\omega^+ \to j j$. We assume that $m_\eta>m_T$, so that $\eta$ plays no role in the top partner decays. We focus on the decay $T\to b\omega^+$, which we assume is the dominant top partner decay. This assumption is a good approximation for $m_T> m_\omega \gsim m_T-m_t$. If $m_\omega < m_T-m_t$, the top partner would decay in both $b\omega^+$ and $t\omega^0$ channels. The latter channel produces signals identical to the ones considered in \Sec{sec:recast1} above. Since we will find that the mass bounds on the $b\omega^+$ and $t\omega^0$ channels are quite similar, we do not attempt a detailed combination of the two; either one can be taken as a good estimate of the bound on this scenario in the region $m_\omega < m_T-m_t$.

\subsubsection{Exact t-Parity}
\label{sec:bMET}

In this case, $\omega^0$ escapes the detector undetected, resulting in a $2b+\met$ signature. The signal topology is identical to sbottom squark ($\tilde{b}$) pair-production, where the sbottom decays via $\tilde{b} \to b \tilde{N}$ and $\tilde{N}$ is a stable neutralino The strongest bounds can be derived from the ATLAS search for two b-jets and missing transverse momentum~\cite{Aad:2013ija}. We recast this search in terms of our signal model using {\tt CheckMate}. The 95\% C.L. constraints on the $m_{T}$-$m_{\eta}$ plane are shown in the left panel of Fig.~\ref{fig:recast2}. For light $\omega$, the top partner masses up to at least 800 GeV are ruled out; the true bound is probably higher, but  no information on cross section bounds beyond 800 GeV was provided in~\cite{Aad:2013ija}. Again, the bound is weakened significantly if $T$ and $\omega$ are quasi-degenerate, even for a rather modest degree of degeneracy: for example, $m_T=500$ GeV is allowed if $m_T-m_\omega \lsim 100$ GeV.

In the compressed region, we again evaluate constraints from the mono-jet search~\cite{Aad:2014nra}, recasting it using {\tt CheckMate}. The excluded region is shown in the left panel of Fig.~\ref{fig:recast2} (purple line). We conclude that top partners as light as 400 GeV are allowed, as long as $T$ and $\omega$ are degenerate at a ${\cal O}(10\%)$ level.

\subsubsection{Approximate t-Parity}
\label{sec:compressed}

The decay chain of interest here is $T \rightarrow b (\omega^+ \rightarrow j j )$. Notice that the $T$ pair production signature here closely resembles the gluino pair production signal, with R-parity violating decay $\tilde{g} \rightarrow b j j$. Thus the search recast in Section~\ref{sec:topsjets} is relevant also in this case. We proceed as before using the same search category. The results are shown in the right panel of Fig.~\ref{fig:recast2}. For generic spectra, the top partner mass below $700-750$ GeV is excluded. The near-degenerate region, $m_T\approx m_\omega$, is not constrained by this search: the two b's present in the final state are required to pass a $p_T > 80$ GeV cut and will often fail in this region. In this case, the signal topology is similar to a pair of massive particles decaying into two jets each. With this in mind we recast a dedicated CMS search looking for pair-produced dijet resonances~\cite{Khachatryan:2014lpa}. This search is also motivated by RPV supersymmetry and is specifically intended for stop pair production with RPV decays in two (light) quarks. Events with at least 4 jets with $p_T > 80$ GeV or $p_T > 120$ GeV and $|\eta|<2.5$ are selected, and a series of additional cuts pair the four leading jets in order to reconstruct two objects with invariant mass close to each other. Given the large background, the signal is searched for as a bump on top of a continuous distribution of events with variable dijet invariant mass. To recast the analysis, we fix $m_T$ and $m_\eta$, simulate the original $\tilde{t}\tilde{t}$ signal for corresponding $m_{\tilde{t}}$ and $m_{\tilde{N}}$ and compute the cut efficiencies, and repeat the procedure for the $T\bar{T}$ signal. Finally we rescale the $T$ pair production cross section by the ratio of the efficiencies,  and extract a limit on $m_T$ corresponding to the $95\%$~C.L. upper bound of the CMS analysis. The result is shown in Figure~\ref{fig:recast2}, as the red shaded area on the right panel. The results are consistent with CMS bounds once the difference in the pair production cross section between fermionic and scalar top partners is taken into account, which amounts to a factor of $\sim 6$. We conclude that top partner masses below about $550$ GeV are excluded for any value of $m_\omega$.

We conclude this section by noticing that CMS has performed a similar search for pair production of 3-jet resonances~\cite{Chatrchyan:2013gia}. This search places gluino mass bounds very similar to those of the ATLAS multi jet search recast above, and the limitations of the two searches, such as the jet $p_T$ cuts that degrade the efficiency in the $m_\omega \approx m_T$ region, are also similar. Thus, we do not expect the CMS search to add significantly to the recast bounds from the ATLAS search shown in Figs.~\ref{fig:recast1} and~\ref{fig:recast2}.

\subsection{Scenario 3: Cascade Top Partner Decays}

In this scenario, the parameter space is mode complicated than in the other two, since $m_T$, $m_\eta$ and $m_\omega$ all play a role. However, the signal topologies are quite similar. The typical final states are $2b+4j+\met$ (exact t-parity) or $2b+8j$ (approximate t-parity), the same as in the scenario 1, $T\to t\eta$, with hadronic top decays. There may be slight differences in the kinematic distributions since no on-shell tops/$W$s are present in the cascade scenario, but we do not expect them to have a significant effect on the mass bounds. The only possibility to significantly relax the top partner mass bound seems to be to assume an approximate triple mass degeneracy of $m_T$, $m_\eta$ and $m_\omega$, and exact t-parity. This case is very similar to the decay $T\to t\eta$ with an off-shell top, which was already considered in Scenario 1.

\section{The Oddest Littlest Higgs}
\label{sec:OddestModel}

In this section we finally describe a non-linear sigma model which can reproduce the simplified model used in earlier sections in certain regions of its parameter space. We use the Littlest Higgs coset~\cite{ArkaniHamed:2002qy}, SU(5)/SO(5), which preserves custodial symmetry and provides a collective tree level quartic for the Higgs. The Goldstones are parameterized by the field $\Sigma$
\begin{equation}
\Sigma = e^{i \Pi_\text{odd} / f}~e^{i \Pi_h /f}~\Sigma_0~e^{i \Pi_h^T /f}~e^{i \Pi_\text{odd}^T/f}~=~e^{i \Pi_\text{odd} / f}~e^{2 i \Pi_h /f}~e^{i \Pi_\text{odd} / f}~\Sigma_0,\\
\end{equation}
with
\begin{equation}
\Sigma_0 = \begin{pmatrix} & & \mathbf{1}_2 \\ & 1 & \\ \mathbf{1}_2 & & \end{pmatrix},\\
\end{equation}
and we have chosen to separate the Goldstone fields as follows:
\begin{align}
\Pi_\text{odd} &= \begin{pmatrix} \omega - \eta/\sqrt{20} &0 & \phi \\ 0 & \sqrt{8/10} \eta & 0 \\ \phi^{\dagger} & 0& \omega^T - \eta/\sqrt{20} \end{pmatrix}, ~~~~~~ \Pi_h =  \begin{pmatrix}0 & H^*/\sqrt{2} & 0\\ H^T/\sqrt{2} &0 & H^\dagger/\sqrt{2} \\ 0& H/\sqrt{2} & 0 \end{pmatrix}.
\end{align}
As is typical in Little Higgs models based on this coset, the pNGB $\phi$ will get a quadratically divergent contribution to its mass and is generically expected to be heavier than the other scalars. We impose a t-parity symmetry which has the following action on the scalar sector
\begin{equation}
\Sigma \to \Sigma^t \equiv \Omega_\Sigma \Sigma^\dagger \Omega_\Sigma, ~~~~~  \Omega_\Sigma = \begin{pmatrix} & & \mathbf{1}_2 \\ & -1 & \\ \mathbf{1}_2 & & \end{pmatrix}.
\end{equation}
On the Goldstone fields, this has the action:
\begin{equation}
H \to H, ~~~~~ \eta \to - \eta, ~~~~~ \omega \to - \omega, ~~~~~ \phi \to -\phi.
\end{equation}
In contrast to the original Littlest Higgs construction, we gauge only the SM SU(2)$_L$ and U(1)$_Y$ subgroups of SU(5), with generators:
\begin{equation}
Q^a = \begin{pmatrix}
\sigma^a/2 &&\\
&0&\\
&&-\sigma^{a *}/2
\end{pmatrix}, ~~~~~
Y = \text{diag}(1,1,0,-1,-1)/2.
\end{equation}
The t-parity acts trivially on the gauge fields. The $H$ field has the quantum numbers of the SM Higgs, while the Goldstone fields $\eta$, $\omega$, $\phi$ have quantum numbers ${\mathbf 1}_0$, ${\mathbf 3}_0$, ${\mathbf 3}_1$ under SU(2)$_L \times$ U(1)$_Y$. The global symmetry is broken explicitly by the gauge couplings, and also by the Yukawa couplings described below. Quantum effects will then generate a potential for the Goldstone fields. This potential is discussed in detail in Section~\ref{sec:ScalarPot}. For reasonable choices of model parameters, a tachyonic mass term is generated for $H$, triggering EWSB, while all other Goldstones acquire positive mass. It can be easily shown that
\begin{gather}
e^{2 i \Pi_h /f} = \begin{pmatrix}
1&0&0&0&0\\
0&\frac{1}{2}+\frac{1}{2}\sqrt{1-s_h^2}&\frac{i}{\sqrt{2}}s_h&0&-\frac{1}{2}+\frac{1}{2}\sqrt{1-s_h^2}\\
0&\frac{i}{\sqrt{2}}s_h&\sqrt{1-s_h^2}&0&\frac{i}{\sqrt{2}}s_h\\
0&0&0&1&0\\
0&-\frac{1}{2}+\frac{1}{2}\sqrt{1-s_h^2}&\frac{i}{\sqrt{2}}s_h&0&\frac{1}{2}+\frac{1}{2}\sqrt{1-s_h^2}
\end{pmatrix},~~~~~~
s_h \equiv \sin\left(\frac{\sqrt{2}h}{f}\right), \label{equ:sDefinition}
\end{gather}
where $H=(\pi^+, (h+i\pi)/\sqrt{2})^T$ and we dropped the $\pi$ fields that are eaten in the EWSB. Reproducing the $W$ mass requires
\begin{equation}
\left<s_h \right>^2 = 2 \frac{v^2}{f^2}\left( 1- \frac{v^2}{2 f^2} \right)\label{equ:svev},
\end{equation}
with $v =(\sqrt{2} G_F)^{-1/2} \approx  \text{246} \; \text{GeV}$. After EWSB, the t-odd pseudo-Goldstones decompose as
\begin{align}
\omega &= \omega_a \sigma_a/2 = \begin{pmatrix}\omega^0/2 & \omega^+/\sqrt{2} \\ \omega^-/\sqrt{2} & -\omega^0/2 \end{pmatrix},\\
\phi &= \phi_a \sigma_a = \begin{pmatrix}\phi^{++} & \phi^+/\sqrt{2} \\ \phi^+/\sqrt{2} & (-\phi^0 + i \phi^0_P)/\sqrt{2} \end{pmatrix}.
\end{align}

In order to build a Lagrangian of the form of Eq. (\ref{equ:schematiclagrangian}), a candidate operator $\mathcal{O}$ can be added in the following way:
\begin{equation}
\mathcal{L} \supset  \left( \mathcal{O} + \mathcal{O}^t \right) + \epsilon \left( \mathcal{O} - \mathcal{O}^t \right),
\end{equation}
where $\mathcal{O}^t$ is the t-image of the operator $\mathcal{O}$, and $\epsilon$ is a small parameter. The top sector of our model consists of a triplet $\chi$, and two singlets $u_1^c$, $u_2^c$ (where the superscript $c$ indicates the field is a color antifundamental, and all fermion fields are two-component left-handed Weyl spinors). The action of t-parity on these fermions is:
\begin{equation}
u_1^c \overset{t}{\longleftrightarrow} u_2^c, ~~~~~ \chi \overset{t}{\longleftrightarrow} \Omega_\chi \chi, ~~~~~ \Omega_\chi = \text{diag}\left(1,1,-1\right).
\end{equation}
The third, odd component of $\chi$ will marry the odd linear combination of $u^c_1$, $u^c_2$, gaining a large Dirac mass and leaving the SM third generation quarks massless before EWSB. The t-preserving top Yukawas are given by:
\begin{equation}\label{equ:Lyuk}
\mathcal{L}_\text{Yuk} = -\frac{y_t}{4} f \left(\chi^i \mathcal{O}_i u^c_1 +\chi^{t i} \mathcal{O}^t_i u^c_2\right)  + \text{h.c.},
\end{equation}
where
\begin{align}\label{equ:Ooperators}
\mathcal{O}_i &= \epsilon_{i j k} \epsilon_{ x y} \Sigma_{j x} \Sigma_{ k y};\\
\mathcal{O}^t_i &= \epsilon_{i j k} \epsilon_{ x y} \Sigma^t_{j x} \Sigma^t_{ k y}.\notag
\end{align}
Here all repeated indices are summed over: $i,j,k = 1,2,3$ and $x,y = 4,5$.
The Higgs is protected by two SU(3) subgroups of the full SU(5), and these are interchanged by t-parity. Each term in this Lagrangian breaks one SU(3) while preserving the other, so the full global symmetry protecting the Higgs is completely broken only non-locally in theory space. This guarantees the absence of quadratic divergences in the Higgs mass$^2$ at one-loop, ameliorating the little hierarchy problem~\cite{ArkaniHamed:2002qy}.

In the top sector, the mass eigenbasis before EWSB is obtained by the following field redefinitions:
\begin{align}
t^c &= \frac{1}{\sqrt{2}}\left(u^c_1 + u^c_2 \right),\\
T^c &=  \frac{1}{\sqrt{2}}\left(u^c_1 - u^c_2 \right).\\
\chi &= \begin{pmatrix}\sigma_2 \cdot Q, T \end{pmatrix}.
\end{align}
Expanding out the $\Sigma$ field to quadratic order in $H$, the Lagrangian reads:
\begin{align}\label{equ:LHiggs}
\mathcal{L}_\text{Higgs} &= - \frac{y_t}{\sqrt{2}}f T T^c + y_t H Q  t^c + \frac{y_t}{\sqrt{2}}\frac{ |H|^2}{f} T T^c+ \text{h.c.}
\end{align}
It can be easily seen that the quadratic divergence from the $T$-loop cancels that of the $t$-loop by noticing that the trace of the Higgs dependent masses, $\Tr M^2(h) = m_T^2(h)+m_t^2(h)$, vanishes at order $h^2$. Before electroweak symmetry breaking the odd top partner $T$ gets a mass $\hat{m}_T \equiv y_t f/\sqrt{2}$, and the top quark is massless. After EWSB, the leading couplings of the 3rd generation quarks to the Goldstones is given by:
\begin{align}
\mathcal{L} &\supset   \frac{1}{2}y_t f s_h t_L t^c + \sqrt{\frac{2}{5}} i y_t T t^c \eta + \frac{i y_t}{2 \sqrt{2}} s_h b_L T^c \left(\omega^- - \phi^- \right)  \\
&~~~~~~ \frac{i y_t}{2\sqrt{2}}s_h t_L T^c \left(\frac{1}{\sqrt{10}}\eta + \frac{1}{\sqrt{2}} \omega^0 - \phi^0 - i \phi^0_P \right) + \text{h.c.}, \notag
\end{align}
where $s_h$ is defined in Eqs. (\ref{equ:sDefinition}), (\ref{equ:svev}). These are exactly the t-preserving couplings of  Eq. (\ref{equ:simplag}). It can be seen that the leading decay for the top partner will be $T \to t \eta$ if this channel is kinematically available, as the decays to $\phi$ and $\omega$ involve couplings suppressed by $\left< s_h\right> \sim v/f$.
However, if the mass splitting between $T$ and $\eta$ is sufficiently small so that this decay cannot proceed on-shell then the decays $T \to b \omega^+$ may dominate if either of these are kinematically available. We note that while the doubly charged scalar $\phi^{\pm\pm}$ could result in some striking signatures, it is unlikely to play an important role in the phenomenology of the top partner due to its quadradically divergent mass and due to the fact that its couplings to the top partner only arise at higher order in the $v/f$ expansion.

In the phenomenological analysis of Sections 2 and 3, we also considered a scenario with approximate t-parity, where the pseudo-Goldstones $\eta$ and $\omega$ may decay to quark pairs. To incorporate this possibility in the OLH model, we can introduce couplings of the form
\begin{equation}
\mathcal{L}_\text{odd} \supset Q^{\hat{i}}_a \boldsymbol{\epsilon}^{(u)}_{ab} u^c_b \left(\mathcal{O}_{\hat{i}} - \mathcal{O}^t_{\hat{i}} \right) + Q^{\hat{i}}_a \boldsymbol{\epsilon}^{(d)}_{ab} d^c_b \left(\mathcal{O}^*_{\hat{i}} - \mathcal{O}^{*t}_{\hat{i}} \right),
\end{equation}
where all repeated indices are summed over: $\hat{i}=1, 2$ while $a, b$ run over three generations. The flavor structure of the $\boldsymbol{\epsilon}$ couplings will determine the decays of the lightest t-odd state.

\subsection{The Scalar Potential}
\label{sec:ScalarPot}

In this section we describe qualitatively the contributions to the Goldstone potential, leaving the lengthy explicit formula to the Appendix. We introduce a tree level mass for the Goldstones by including the following explicit global symmetry breaking (but custodial and t-parity invariant) term in to the scalar potential:
\begin{equation}
V \supset f^2 \text{Tr} \left[ M \Sigma\right] + \text{h.c.}
\end{equation}
where:
\begin{equation}
M=\frac{1}{32}\begin{pmatrix}
&&4 m_2^2\\
&5 m_1^2 - m_2^2&\\
4 m_2^2&&
\end{pmatrix}.
\end{equation}
This particular normalization is chosen for convenience after expanding out the $\Sigma$ field in terms of the Goldstones. When expanded in terms of the Goldstone fields, it introduces a mass for $\eta$ of $m_1$, and a mass contribution for $\omega$ and $\phi$ of $m_2$. In order to reproduce the compressed spectrum of section \ref{sec:compressed}, we will need to make $m_\eta >  m_T - m_t$. This will require us to explore the region of parameter space where $m_1^2$, and possibly also $m_2^2$ are not negligibly small compared to $f$. A precise study would require considering all operators that can be constructed, consistent with the symmetries, in powers of $M/f^2$. However for the purposes of this work, we only introduce the additional operators that will add qualitatively new features to the potential, setting the other coefficients to zero for simplicity.

Quadratically divergent fermion loops involving the couplings in Eqs. (\ref{equ:Lyuk}), require the introduction of a counterterm:
\begin{equation}\label{equ:cTterm}
\mathcal{L} \supset \frac{y_t^2}{8} f^4 c_T \epsilon_{i j k} \epsilon_{k l m} \epsilon_{x y} \epsilon_{w z} \Sigma_{i x} \Sigma_{j y} \Sigma^*_{l w} \Sigma^*_{m z} + \text{t-image},
\end{equation}
where $c_T$ is an $\mathcal{O}(1)$ number determined by UV physics. This contributes the ordinary tree level collective quartic for the Higgs, as well as a large contribution to the $\phi$ mass. We also include the operator:
\begin{equation}
\mathcal{L} \supset \frac{y_t^2}{4} f^2 c_{TM} \epsilon_{i j k} \epsilon_{k l m} \epsilon_{x y} \epsilon_{w z} M_{i x} \Sigma_{j y} \Sigma^*_{l w} \Sigma^*_{m z} + \text{t-image} + \text{h.c.}.
\end{equation}
which also typically has an $\mathcal{O}(1)$ coefficient $c_{TM}$ and a parametric suppression $M/f^2$. This operator contributes to the masses of the all of the Goldstones.

The additional fermion loop contributions to the Goldstone potential are calculated using the Coleman-Weinberg (CW) potential \cite{PhysRevD.7.1888}
\begin{equation}
V_\text{CW} = -\frac{N_c}{32 \pi^2}\sum_i M_i^4 \left(\log\left(\frac{M_i^2}{\Lambda^2}\right) - \frac{3}{2} \right)
\end{equation}
where the sum is over the eigenvalues of the fermion mass matrix. There is a log-divergent piece which contributes to the Higgs quartic and $\phi$ mass which is degenerate with the quadratic divergence in Eq. (\ref{equ:cTterm}) and can therefore be absorbed by a redefinition of $c_T$. Remaining log divergences are cut off at a scale $\Lambda = y_2 f$, with $y_2 \sim \mathcal{O}(2)$. This may be the scale of new fermion resonances, an example of which is given in Appendix (\ref{app:morefermions}). This log divergent and additional finite parts contribute to both the Higgs mass and quartic, but only contributes to the masses of the other Goldstones after EWSB and so this effect is suppressed by $v^2/f^2$.

Quadratically divergent gauge boson loops require counterterms of the form
\begin{equation}
\mathcal{L} \supset c_2 g_2^2 f^4 \text{Tr}\left[ Q \Sigma Q^* \Sigma^* \right] + c_Y g_Y^2 f^4 \text{Tr}\left[Y\Sigma Y \Sigma^* \right].
\end{equation}
These operators provide tree level contributions to $m_H^2$, $m_\phi^2$, $m_\omega^2$, and the Higgs quartic, and sub-leading corrections and mixings after EWSB. Additional terms obtained by including insertions of the mass matrix $M$ are degenerate with a redefinition of the mass matrix.

For obtaining the correct Higgs potential in the compressed scenario, we also introduce the following term which explicitly breaks all of the symmetries protecting the Higgs
\begin{equation}
\mathcal{L} \supset \frac{5}{128} f^2 m_3^2 \left(\Sigma_{33}^2 + \text{h.c.}\right).
\end{equation}
This operator provides positive masses for $\eta$ and $h$, but does not contribute to the Higgs quartic. The role of this term will be discussed in more detail in Appendix (\ref{sec:OLHpot}).

\subsection{Sample Spectra}
The top partner has mass $\mathcal{O}(f/\sqrt{2})$. The pNGBs $\phi$ and $\omega$ get quadratically divergent contributions to the masses, typically raising them significantly above the Higgs mass unless there is some additional tuning. On the other hand, the loop generated mass for $\eta$ is of order $v/\sqrt{2}\pi$ and so unless there are large tree level contributions to its mass it tends to be somewhat lighter than the Higgs.

In tables (\ref{tab:input}), (\ref{tab:output}), we show sample parameter space points of the Oddest Littlest Higgs which reproduce the simplified phenomenological models of section (\ref{sec:simpmodels}). Case A1 is typical if the tree level breakings of the global symmetry are small, with $\eta$ being the lightest pNGB. The decay $T \to t \eta$ dominates and so this reproduces scenario 1 of \Sec{sec:simpmodels}, with the $\eta$ decaying into two hard jets. Case A2 has a very light $\eta$ such that it will be highly boosted when produced from decays of $T$, and so its decay products will be observed as a single jet. This places it in the narrow window of \Fig{fig:recast1} for light $\eta$ where the exclusion limits are weaker, but the model parameters are tuned to avoid a large branching fraction $h \to \eta \eta$. Case B has a compressed spectrum, with the top partner decaying via $T \to b\omega^+$ as the decay $T \to t\eta$ is not kinematically available. Raising the $\eta$ mass is achieved via large tree level contributions from $m_1$ and $m_3$. In this scenario, the dominant contributions to the tuning in the Higgs mass parameter are actually coming from $m_1$ and $c_Y$, with top loops being subleading. A naive estimate of the tuning in the Higgs mass parameter coming from these contributions is $\mathcal{O}(5\%)$, as discussed in Appendix (\ref{sec:OLHpot}). In case C, the mass hierarchy will lead to a cascade decay of the form $T \to b \omega^+ \to b q \bar{q} \eta$. This possibility was mentioned in \Sec{sec:simpmodels}, although we have not discussed it in detail.

\begin{table}
\centering
\begin{tabular}{l|l l l l l l l l l}
Case&$f/\text{GeV}$  &$(m_1/\text{GeV})^2$&$(m_2/\text{GeV})^2$&$(m_3/\text{GeV})^2$&$ y_2$&$ c_T$&$ c_{T M}$&$ c_Y$&$ c_2$\\ \hline
A1     & 1320                & $200^2$               & $100^2$                 &0                         & 2        & 0.07   & 1.0      & -0.1      &0.013\\
A2    & 1150               & $155^2$                  & 0                            & $-160^2$              & 1.5    & 0.084  &0      & -0.3         & 0.039\\
B     & 890                 & $-295^2$                  & $200^2$                 &  $690^2$               & 2      & 0.29   & 2.5      & -2.05    & 0.091\\
C     & 890                & $-310^2$                   & $200^2$                  &$630^2$                & 2        & 0.25   &3          & -1.77   & 0.088
\end{tabular}
\caption{Input Lagrangian parameters for sample spectra.}
\label{tab:input}
\end{table}

\begin{table}
\centering
\begin{tabular}{l|l l l}
Case&$ m_T/\text{GeV}$&$\left\{m_{\phi^0}, m_{\omega^0},m_\eta \right\}/\text{GeV}$&$ \left\{m_{\phi^\pm},m_{\omega^{\pm}} \right\}/\text{GeV}$\\ \hline
A1     & 900  & \{600, 300, 200\}                     & \{600, 300\}       \\
A2     & 810 & \{560, 400, 33\}                        & \{570, 400\}       \\
B     & 600   & \{590, 560, 580\}                    & \{590, 560\}       \\
C     & 600 & \{600, 560, 510\}                      & \{610, 560\}
\end{tabular}
\caption{Sample mass hierarchies.}
\label{tab:output}
\end{table}

\section{Conclusions and Outlook}
\label{sec:conc}

Fermionic top partners are well motivated theoretically, and form an important component of new physics search program at the LHC. Currently, the experimental searches focus on three decay topologies: $T\to bW$, $T\to tZ$, and $T\to th$. However, top partners may carry new conserved quantum numbers that forbid these decays. The simplest possibility is a conserved parity, under which the top partner is odd and all SM states are even. In this case, decays of top partners may involve new particle-odd scalars, leading to non-standard experimental signatures. If the parity is exact, the lightest particle-odd scalar is stable, and assuming that it is weakly interacting, the scenario is characterized by missing transverse energy signatures, with signal topologies identical to stops in R-parity conserving supersymmetry. If, on the other hand, the parity is only approximate, the lightest parity-odd scalar may decay, for example, into jets, resulting in multi-jet or tops+jets final states similar to those produced by gluinos and stops in R-parity violating supersymmetry. In either case, we found that the current LHC lower bounds on the top partner mass are similar to those in the conventional decay scenario, $m_T\gsim 700-900$ GeV, if the mass of the lightest t-odd scalar is well below $m_T$. If, on the other hand, the top partner and the lightest t-odd scalar are somewhat degenerate in mass, the bounds can be relaxed significantly. For example, in the case of exact t-parity and decays into a gauge-singlet scalar $\eta$, a 500 GeV top partner is allowed as long as $m_\eta$ is between 325 and 500 GeV. The low allowed top partner mass reduces the need for fine-tuning in the Higgs mass parameter, compared to the conventional decay scenario, making this class of models a theoretically attractive possibility. In the OLH model considered in~\Sec{sec:OddestModel}, this can only be achieved at the expense of introducing new tunings of tree-level parameters associated with raising the mass of $\eta$. It remains an interesting open question whether a similar model can be constructed in which a compressed spectrum can be arranged without directly impacting the tuning of the Higgs mass parameter.  

An interesting issue not investigated here is the possibility that the t-parity is anomalous~\cite{Hill:2007nz,Hill:2007zv}. Whether or not such an anomaly is present depends entirely on the UV completion of the TeV-scale NLSM~\cite{Krohn:2008ye,Csaki:2008se}, and it is certainly consistent to assume that the anomaly is absent. If it were present, it would give rise to a phenomenologically interesting possibility of the lightest t-odd scalar decaying to two SM massive vector bosons, for example $\eta\to ZZ$. Depending on the size of the explicit t-parity violating couplings, these decays may become dominant. Hadronic $Z$ decays would give rise to signatures similar to the ones considered in the approximate t-parity scenarios we studied, but with higher jet multiplicity and softer jets. Leptonic $Z$ decays may also be exploited in this case. We leave a detailed analysis of this possibility for future work. 

A natural by-product of our scenario is that, if the t-parity is exact and non-anomalous, the lightest t-odd particle can be a dark matter candidate. Unlike the LHT models, where the stable dark matter candidate is usually a spin-1 T-odd partner of the hyper charge gauge boson~\cite{Birkedal:2006fz}, in this case the dark matter particle would be a scalar. It would be interesting to understand if the correct relic abundance can be obtained in viable and phenomenologically interesting regions of the model parameter space. 

\section*{Note Added:}
While we were completing this manuscript, we became aware of Ref.~\cite{Serra:2015} where similar ideas were pursued in the context of holographic Composite Higgs models.

\section*{Acknowledgements}
We are grateful to James Alexander, Gustaaf Brooijmans, Jeff Dror, Nathan Mirman, Javier Serra, and Ennio Salvioni for useful discussions related to this work. This research is supported by the U.S. National Science Foundation through grant PHY-1316222. We also acknowledge the support of the Bethe Postdoctoral Fellowship (EK) and the John and David Boochever Prize Fellowship in Fundamental Theoretical Physics (JC).

\appendix
\section{An Extended Fermion Sector for the Oddest Littlest Higgs}\label{app:morefermions}

In this section we describe an extended fermion sector for the OLH model which cuts off log divergences in the Higgs potential that are not degenerate with the quadratically divergent piece responsible for the collective quartic. The top sector of this model consists of three triplets $\chi_1$, $\chi_2$, $\chi^c$, and two singlets $u_1^c$, $u_2^c$. The action of t-parity on these fermions is:
\begin{equation}
\chi_1 \overset{t}{\longleftrightarrow} \chi_2, ~~~~~ u_1^c \overset{t}{\longleftrightarrow} u_2^c, ~~~~~ \chi^c \overset{t}{\longleftrightarrow} \Omega_\chi \chi^c, ~~~~~ \Omega_\chi = \text{diag}\left(-1,-1,1\right).
\end{equation}
We will see that these fields decompose in to a t-even SM left handed doublet and right handed singlet (the top quark and left-handed bottom), a light t-odd singlet top partner which cancels the quadratic divergence of the top, and a heavy triplet of fermions -- an odd doublet, and an even singlet. The t-preserving top Yukawas are given by:
\begin{equation}
\mathcal{L}_\text{Yuk} = -\frac{y_1}{2} f \left[\chi_1^i \mathcal{O}_i u^c_1 +\chi_2^i \mathcal{O}^t_i u^c_2\right] + \frac{y_2}{\sqrt{2}} f \left[(\chi_1 \cdot \chi^c + \chi_2 \cdot {\Omega_\chi} \cdot \chi^c) \right] + \text{h.c.}, ~~~ i = 1,2,3, \label{equ:Lyuk2}
\end{equation}
where $\mathcal{O}_i$ and $\mathcal{O}^t_i$ are given as in Eq. (\ref{equ:Ooperators}). These Yukawas are very similar to those in \cite{Cheng:2005as}, except that because t-parity acts trivially on the gauge sector, we assume that the fermion multiplets transform as incomplete linear representations of the SU(5) global symmetry group and we do not require that they have non-linear transformations under SO(5)\footnote{An extension of the gauge sector at $\sim$ (few TeV) would require that t-parity act non-trivially on the full gauge group, necessitating the introduction of complete multiplets or non-linear symmetry transformations on the fermions.}.

The triplets decompose in the following way:
\begin{equation}
\chi_1 = \frac{1}{\sqrt{2}}\begin{pmatrix}Q + Q'\\ T + T' \end{pmatrix},~~~~~~~~
\chi_2 = \frac{1}{\sqrt{2}}\begin{pmatrix}Q - Q'\\ -T + T'\end{pmatrix},~~~~~~~~
\chi^c = \begin{pmatrix}{Q'}^c \\u^c_3 \end{pmatrix}
\end{equation}
and then we make the following field redefinitions:
\begin{align}
t^c &= \frac{1}{\sqrt{y_1^2+y_2^2}}\left( y_1 u^c_3 - \frac{y_2}{\sqrt{2}} \left(u^c_1 + u^c_2\right) \right),\\
{T'}^c &= \frac{1}{\sqrt{y_1^2+y_2^2}}\left( y_2 u^c_3 + \frac{y_1}{\sqrt{2}} \left(u^c_1 + u^c_2\right) \right),\\
T^c &=  \frac{1}{\sqrt{2}}\left(u^c_1 - u^c_2 \right).
\end{align}
Expanding out the $\Sigma$ field to leading order in $H$, the Lagrangian takes a particularly simple form in this new basis:
\begin{align}\label{equ:Lleading}
\mathcal{L}_\text{leading} &= - \frac{y_2 y_t}{\sqrt{2 y_2^2 - y_t^2}}f T T^c - \frac{\sqrt{2}y_2^2}{\sqrt{2 y_2^2 - y_t^2}}f T' {T^c}' +  f y_2 Q' {Q^c}'\\
&~~~~~~~~ + y_t (Q H) t^c +  \frac{y_t^2}{\sqrt{2 y_2^2 - y_t^2}}(QH){T^c}' + \frac{\sqrt{2} y_2 y_t}{\sqrt{2 y_2^2 - y_t^2}}(Q' H) T^c + \text{h.c.}, \notag
\end{align}
where we have replaced $y_1$ by $y_t$:
\begin{equation}
y_t^2 = \frac{2 y_1^2 y_2^2}{y_1^2 + y_2^2}.
\end{equation}
In the limit $y_2 \gg y_t$, Eq. (\ref{equ:Lleading}) reduces to:
\begin{align}\label{equ:Lleading2}
\mathcal{L} &\supset - \frac{ y_t}{\sqrt{2}}f T T^c - y_2 f\left( T' {T^c}' +   Q' {Q^c}'\right)+ y_t (Q H) t^c +  \frac{y_t^2}{\sqrt{2} y_2}(QH){T^c}' + y_2(Q' H) T^c.
\end{align}
This limit is a decoupling limit, in which the primed fields form a heavy and nearly degenerate triplet, leaving just the physical top quark and the odd top partner in the low energy spectrum. The parities of the various fields are:
\begin{center}
\begin{tabular}{c|c}
$+$ & $-$\\ \hline
$Q, t^c, {T^{(c)}}'$ & $T^{(c)}, {Q^{(c)}}'$
\end{tabular}.
\end{center}

Before electroweak symmetry breaking, the primed fields acquire a mass $\hat{m}' \equiv y_2 f$, the odd top partner $T$ gets a mass $\hat{m}_T \equiv y_t f/\sqrt{2}$, and the top quark is massless. Integrating out the primed fields at tree level will generate custodial symmetry violating couplings for the light top partner, which will generate corrections to the T parameter at one-loop. These fields also serve to cut off the logarithmic divergences in the loop generated Higgs potential of the OLH. In the limit $y_2 \gg y_t$, this is the only role they play and can otherwise be ignored in the collider phenomenology of the model.

\section{Oddest Littlest Higgs Potential}
\label{sec:OLHpot}
The Higgs potential is given by:
\begin{equation}\label{equ:spotential}
V_\text{higgs} = \frac{1}{4}m^2 f^2 s_h^2 + \frac{1}{16}\lambda f^4 s_h^4 - \frac{3}{16 \pi^2} m_t^4\left(s_h^2\right)\left(\log\left(\frac{m_t^2(s_h^2)}{\mu^2}\right) - \frac{3}{2}\right) + \mathcal{O}\left(s_h^6 \right)
\end{equation}
where $m_t^2\left(s_h^2\right)$ is the Higgs-dependent top mass:
\begin{equation}
m_t^2 = \frac{1}{4}y_t^2 f^2 s_h^2 + \mathcal{O}\left(s_h^4\right).
\end{equation}
The scale $\mu$ in Eq. (\ref{equ:spotential}) will be set to the top mass so that the log vanishes at the potential minimum, though the term still plays a role in setting the minimum of the potential. The potential is minimized with:
\begin{equation}
m^2 = - \frac{1}{2}\lambda f^2 s_h^2 + \frac{3}{32 \pi^2} y_t^4 f^2 s_h^2
\end{equation}
resulting in a physical Higgs mass:
\begin{align}
m_h^2 = \left(125 \; \text{GeV}\right)^2 &= \lambda f^2 s_h^2 \left(1-s_h^2\right)\\
&= 2 \lambda v^2 \left(1 - \frac{5 v^2}{2 f^2}\right). \notag
\end{align}

\begin{figure}
\captionsetup[subfigure]{labelformat=empty}
\centering
\subfloat[]{\includegraphics[width=0.47\textwidth]{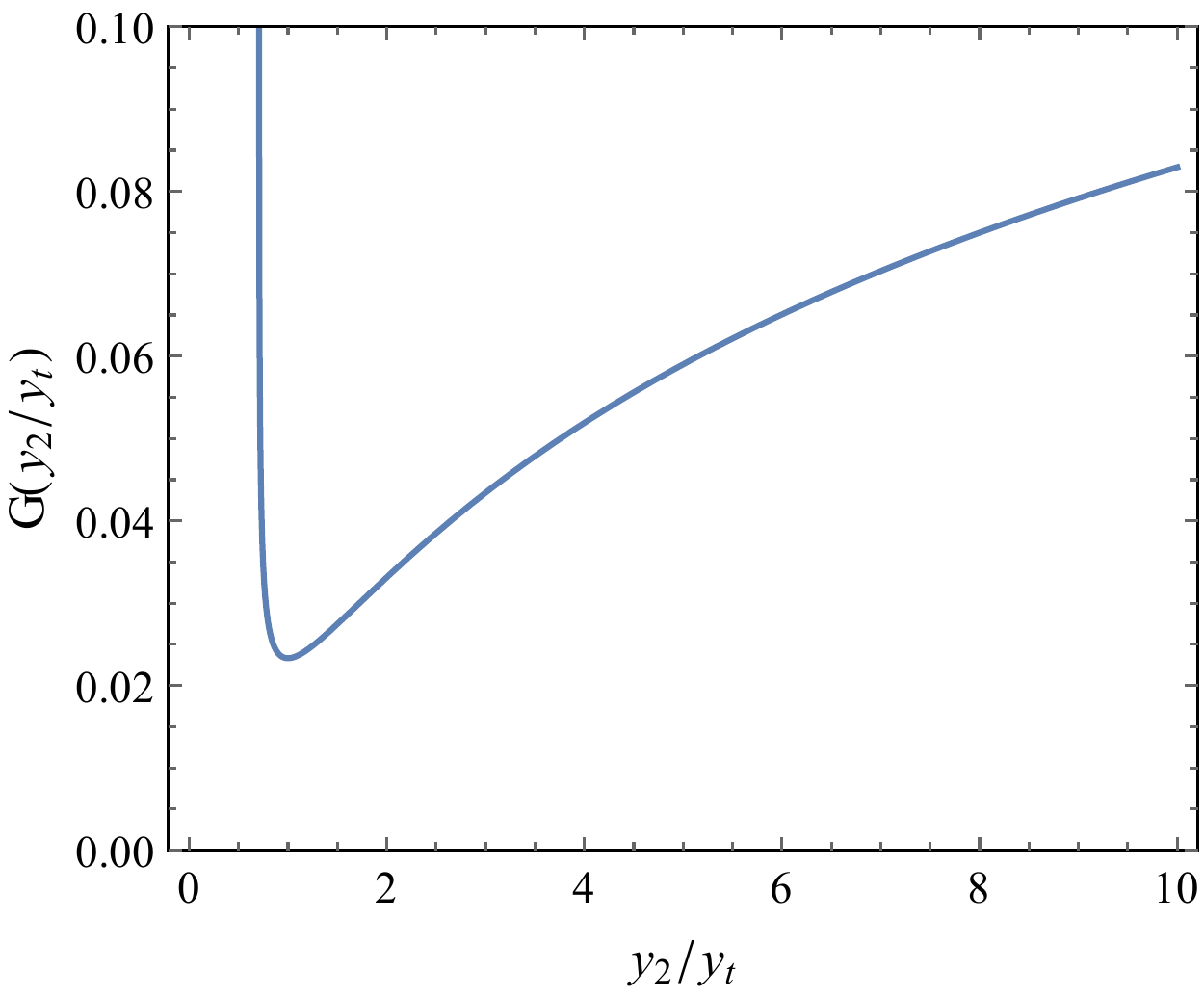}}
\hfill
\subfloat[]{\includegraphics[width=0.47\textwidth]{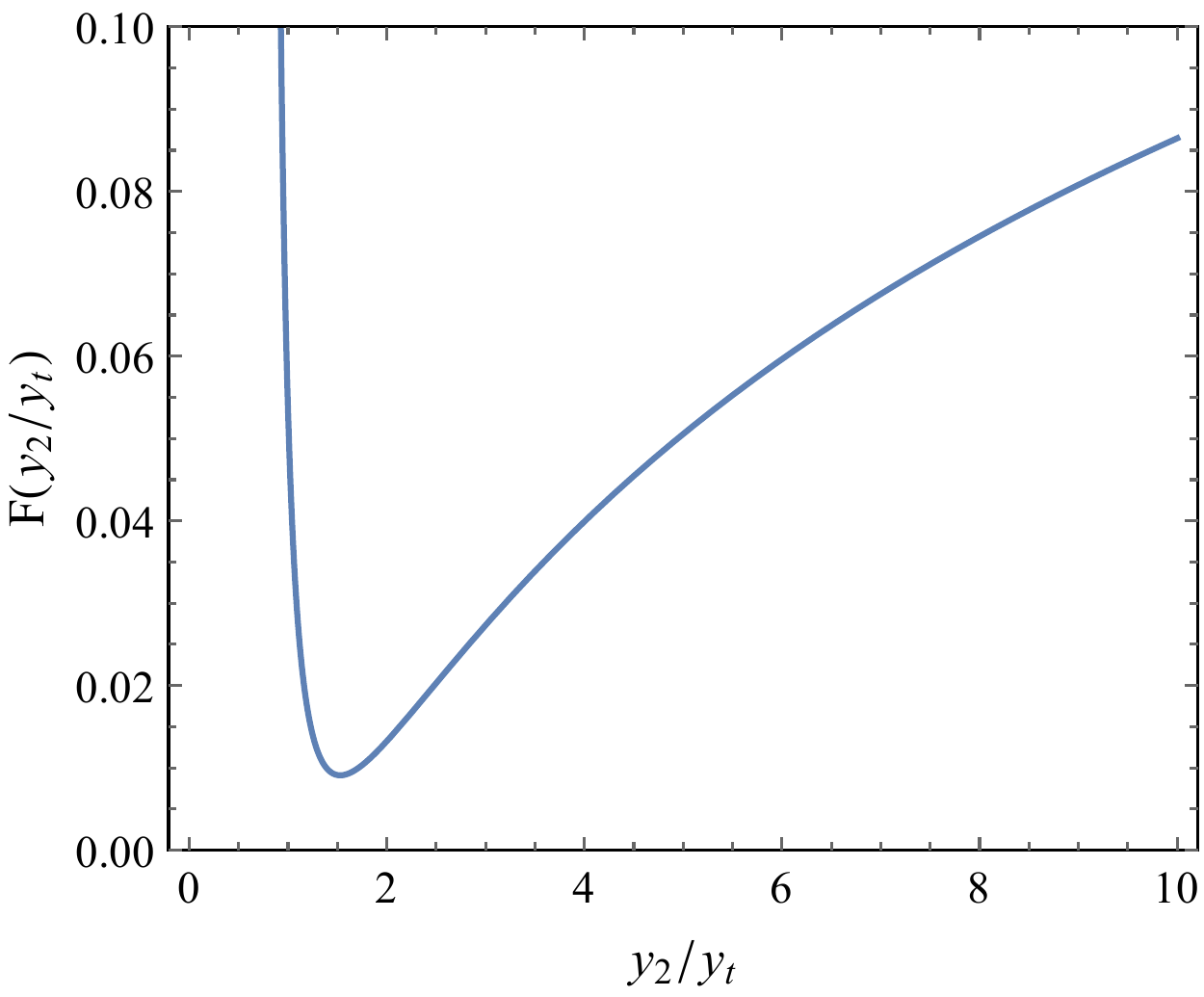}}
\caption{$G$ and $F$ functions which contribute to the Higgs mass parameter and quartic.}
\label{fig:FGfuncs}
\end{figure}

The Goldstones in the Oddest Littlest Higgs model of section (\ref{sec:ScalarPot}) have masses given by (in the limit $y_2 \gg y_t$):
\begin{align}
m_\eta^2 &= m_1^2 + m_3^2 + \frac{1}{10} y_t^2 c_{TM} m_2^2 + \mathcal{O}\left(s_h^2\right),\\
m_\omega^2 &= m_2^2 \left(1 + \frac{1}{2} y_t^2 c_{TM} \right) + 8 c_2 g_2^2 f^2 +\mathcal{O}\left(s_h^2\right),\\
m_\phi^2 &= m_2^2 \left(1 + \frac{3}{2} y_t^2 c_{TM}\right) + 8 c_2 g_2^2 f^2 + 4 c_Y g_Y^2 f^2 + 4 y_t^2 c_T f^2 +\mathcal{O}\left(s_h^2\right).
\end{align}
The mass parameter and quartic of the Higgs potential are given by:
\begin{align}
m^2 &= \frac{3}{8}m_2^2 \left(1 + \frac{2}{3} y_t^2 c_{TM} \right) + \frac{5}{8}m_1^2 + \frac{5}{16}m_3^2 + 3c_2 g_2^2 f^2 + c_Y g_Y^2 f^2 - y_t^4 f^2 G\left(\frac{y_2}{y_t}\right)\label{equ:massparamexpression}\\
\lambda &= \frac{3}{8}\frac{m_2^2}{f^2}\left(1 + \frac{4}{3} y_t^2 c_{TM} \right) + \frac{5}{8} \frac{m_1^2}{f^2} + 3 c_2 g_2^2 + c_Y g_Y^2 + y_t^2 c_T + \frac{3 y_t^4}{16\pi^2} \log \frac{m_T^2}{m_t^2} + y_t^4 F\left(\frac{y_2}{y_t}\right)\label{equ:quarticexpression}
\end{align}
where $F$ and $G$ are contributions generated by fermion loops between the scales of the light and heavy top partners, shown below and plotted in Fig (\ref{fig:FGfuncs}).
\begin{align}
G(x) &= \frac{3}{8 \pi^2}\frac{x^4}{2 x^4 - 3 x^2 + 1}\left[\log \left(2 x^2 \right) + \left(2 x - 1 \right) \log\left(1-\frac{1}{2x^2} \right)\right]\\
F(x) &= \frac{3}{16 \pi^2}\Bigg[-2- \frac{2 x^4}{\left(x^2 - 1 \right)^2} + \log\left(2x^2\right) -\left(1+\frac{4 x^6}{\left(2x^2-1\right)^2}\right) \log\left(1-\frac{1}{2x^2} \right)\\
&~~~~~~~~~~~~~~~~~~ + \left(\frac{x^6}{\left(x^2 - 1 \right)^3} - \frac{2 x^4}{\left(2 x^2 - 1\right)^2} \right)\log\left(2 x^2 - 1 \right) \Bigg]\notag
\end{align}
The expression for the quartic, Eq. (\ref{equ:quarticexpression}), can be rewritten in terms of the Goldstone masses as follows
\begin{equation}
\lambda = \frac{1}{8 f^2}\left(2m_\phi^2 + m_\omega^2 + 5 m_\eta^2 \right) + \frac{3 y_t^4}{16\pi^2} \log \frac{m_T^2}{m_t^2} + y_t^4 F\left(\frac{y_2}{y_t}\right) - \frac{5}{8}\frac{m_3^2}{f^2}.
\end{equation}
In order to arrange for the compressed spectrum of Section (?), it is required that all of the scalars have masses $\gtrsim f/\sqrt{2}$. It is clear from this expression that a small quartic can only be achieved in this case if $m_3$ is large. In case B of the sample spectra, we have obtained a tachyonic Higgs mass parameter and a small quartic using negative $c_Y$ and tachyonic $m_1^2$, which provide important negative contributions in expressions (\ref{equ:massparamexpression}), (\ref{equ:quarticexpression}) to balance the positive contributions that are needed for heavy scalars, while the large $\eta$ mass (required to make the $T \to t \eta$ decays kinematically forbidden) is obtained with a large $m_3$. The cancelation between the contributions from $m_1$ and $c_Y$ is the dominant contribution to the tuning of the Higgs mass parameter in this case. A naive estimate of the tuning in the Higgs mass parameter is given by:
\begin{equation}
\Delta = \left|\frac{\text{Max}\left[\delta m_i^2\right]}{m^2}\right|,
\end{equation}
where $\delta m_i^2$ are the individual contributions to the mass parameter in Eq. (\ref{equ:massparamexpression}). By this measure, case B has a tuning $\Delta^{-1} = 4 \%$. A model which can reproduce the compressed scenario without additional tuning in the Higgs potential would be interesting, and it would require either that the parity-preserving couplings of $\eta$ to the top partner are suppressed, the mass of $\eta$ can be raised without large contributions to the Higgs potential, or that the state does not exist in the first place.

\bibliography{lit}
\bibliographystyle{jhep}

\end{document}